\documentclass[usenatbib,useAMS]{mnras}

\usepackage{soul}
\usepackage{ulem}
\usepackage{longtable}
\usepackage{url}
\usepackage{newtxtext,newtxmath}
\newcommand{\sourceName}{NVSS J141922--083830\xspace}
\newcommand{\NVSS}{NVSS J141922--083830\xspace}

\newcommand{\swift}{{\it Swift}}
\usepackage[T1]{fontenc}
\usepackage{ae,aecompl}
\usepackage{color}

\usepackage{graphicx}	% Including figure files
\usepackage{amsmath}	% Advanced maths commands
\usepackage{nicefrac}
\usepackage{xspace}
\usepackage[super]{nth}
\usepackage[acronym]{glossaries}
\usepackage{multirow}
\usepackage{bbding,pifont}
\usepackage{lineno}
\title[A multiwavelength study of the FSRQ NVSS J141922-083830]{A multiwavelength study of the flat spectrum radio-quasar NVSS J141922-083830 covering four flaring episodes\thanks{based on observations made with the Southern African Large Telescope (SALT)}}

\author[Buckley D. A. H. et al.]{D. A. H. Buckley$^{1,2,3\href{https://orcid.org/00000002-7004-9956}{\includegraphics[scale=0.7]{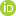}}}$\thanks{E-mail: dibnob@saao.ac.za},
R. J. Britto$^{2,4,\href{https://orcid.org/0000-0003-3456-2362}{\includegraphics[scale=0.7]{ORCIDiD_icon16x16.png}}}$\thanks{E-mail: dr.richard.britto@gmail.com}, 
S. Chandra$^{1,5,\href{https://orcid.org/0000-0002-8776-1835}{\includegraphics[scale=0.7]{ORCIDiD_icon16x16.png}}}$,
V. Krushinsky$^{6}$,
M. B{\"o}ttcher$^{5,\href{https://orcid.org/0000-0002-8434-5692}{\includegraphics[scale=0.7]{ORCIDiD_icon16x16.png}}}$,
\newauthor S. Razzaque$^{4,7,8, \href{https://orcid.org/0000-0002-0130-2460}{\includegraphics[scale=0.7]{ORCIDiD_icon16x16.png}}}$,
V. Lipunov$^{9}$,
C. S. Stalin$^{10\href{https://orcid.org/0000-0002-4998-1861}{\includegraphics[scale=0.7]{ORCIDiD_icon16x16.png}}}$,
E. Gorbovskoy$^{9}$,
N. Tiurina$^{9}$,
D. Vlasenko$^{9}$,
\newauthor and A. Kniazev$^{1,11}$
\\ \\
$^{1}$South African Astronomical Observatory, PO Box 9, Observatory 7935, Cape Town, South Africa\\
$^{2}$Department of Physics, University of the Free State, PO Box 339, Bloemfontein 9300, South Africa\\
$^{3}$Department of Astronomy, University of Cape Town, Private Bag X3, Rondebosch 7701, South Africa\\
$^{4}$Centre for Astro-Particle Physics and Department of Physics, University of Johannesburg, PO Box 524, Auckland Park 2006, South Africa\\
$^{5}$Centre for Space Research, North-West University, Potchefstroom 2520, South Africa\\
$^{6}$Laboratory of Astrochemical Research, Ural Federal University, Ekaterinburg, Russia, ul. Mira d. 19, Yekaterinburg, Russia, 620002\\
$^{7}$Department of Physics, The George Washington University, Washington, DC 20052, USA \\
$^{8}$National Institute for Theoretical and Computational Sciences (NITheCS), South Africa \\ 
$^{9}$Department of Physics, Sternberg Astronomical Institute, Lomonosov Moscow State University, 119234, Universitetsky, 13, Moscow, Russia\\
$^{10}$Indian Institute of Astrophysics, Bangalore, India\\
$^{11}$Southern African Large Telescope, PO Box 9, Observatory 7935, Cape Town, South Africa\\
}

\date{Accepted XXX. Received YYY; in original form ZZZ}

\pubyear{2022}

\begin{document}
%\linenumbers
\label{firstpage}
\pagerange{\pageref{firstpage}--\pageref{lastpage}}
\maketitle

\begin{abstract}
We present multiwavelength observations and a model for flat spectrum radio quasar NVSS J141922-083830, originally classified as a blazar candidate of unknown type (BCU II object) in the Third {\it Fermi}-LAT AGN Catalog (3LAC). Relatively bright flares (>3 magnitudes) were observed on 21 February 2015 (MJD~57074) and 8~September 2018 (MJD~58369) in the optical band with the MASTER Global Robotic Net (MASTER-Net) telescopes.
Optical spectra obtained with the Southern African Large Telescope (SALT) on 1 March 2015 (MJD~57082), during outburst, and on 30 May 2017 (MJD~57903), during quiescence, showed emission lines at 5325\AA{} and at $\approx$3630\AA{} that we identified as the Mg~{\sc ii} 2798\AA{} and C~{\sc iii}] 1909\AA{} lines, respectively, and hence derived a redshift $z$ = 0.903. Analysis of {\it Fermi}-LAT data was performed in the quiescent regime (5 years of data) and during four prominent flaring states in February--April 2014, October--November 2014, February--March 2015 and September 2018. We present spectral and timing analysis with {\it Fermi}-LAT. We report a hardening of the gamma-ray spectrum during the last three flaring periods, with a power-law spectral index $\Gamma = 2.0$--$2.1$. The maximum gamma-ray flux level was observed on 24 October 2014 (MJD~56954) at $(7.57 \pm 1.83) \times 10^{-7}$ ph~cm$^{-2}$s$^{-1}$.
The multi-wavelength spectral energy distribution during the February--March 2015 flare supports the earlier evidence of this blazar to belong to the FSRQ class. The SED can be well represented with a single-zone leptonic model with parameters typical of FSRQs, but also a hadronic origin of the high-energy emission can not be ruled out.%We derived a spectral energy distribution for the source, which corresponds to the one of a flat spectrum radio quasar (FSRQ). The SED can be well represented with a single-zone leptonic model with parameters typical of FSRQs, but also a hadronic origin of the high-energy emission can not be ruled out. At last, we do support the earlier evidence of this blazar to belong to the FSRQ class.
\end{abstract}


%%%%%%%%%%%%%%%%% BODY OF PAPER %%%%%%%%%%%%%%%%%%

\section{Introduction}

An active galactic nucleus (AGN) is a compact structure at the centre of a massive galaxy (usually elliptical). It consists of a super-massive black hole (SMBH, M $\sim$ $10^{6}-10^{10}$ solar masses) surrounded by an accretion disk of plasma that emits thermal radiation $-$ mainly in the UV $-$ produced by the strong gravitational potential of the central SMBH on the plasma particles. Several classes of objects that were discovered or classified throughout the past decades $-$ radio-galaxies, Seyfert galaxies, then quasars and blazars $-$ are today discussed within the framework of the unified model of AGNs \citep{Urry}. AGNs are classified into many classes and subclasses, and they may appear to the observers as objects with different properties, due to their distance, their position with respect to their observation direction, or their physical properties. However, it is most likely that they are all the same type of objects, but viewed through different angles and/or at different stages of their evolution. Quasi stellar radio sources (quasars) and quasi stellar objects (QSOs) are the most distant AGNs, since they are often seen as point-like objects, outshining their host galaxy which is often too faint to be detected. Radio emission from AGNs comes from a pair of Doppler-boosted jets of relativistic plasma, understood to be initiated by a spinning SMBH in a strong magnetic field ($\geq 1$~G). These jets are typically perpendicular to the accretion disk and are pointing in opposite directions. We focus in this paper on the {\it blazar} class, radio-loud AGNs that are seen close to the direction of one of their jets.

Blazars are classified into two subclasses: flat spectrum radio quasars (FSRQs) and BL Lac objects. Most of the radiation detected from blazars is non-thermal emission produced within the jets, and spreads over the whole electromagnetic spectrum, from radio waves to gamma rays. The spectral energy distributions (SEDs) of blazars exhibit a characteristic double bump structure. The low-energy bump peaks in the optical--IR, and is understood to be due to synchrotron emission from relativistic electrons of the jets. The high-energy bump typically peaks in the X-ray or gamma-ray domain. The radiation production mechanisms that causes the high-energy bump are still debated. It could be due to inverse Compton scattering of photon fields by some population of electrons which are producing the synchrotron radiation (leptonic models).
However, in several instances the data may be better reproduced by hadronic models (proton synchrotron, cascades, etc) --- see for example \citet{Boettcher}. Depending on the frequency of the first peak (also called synchrotron peak $\nu_{pk}^{syn}$), we refer to low, intermediate and high synchrotron peaked objects respectively: LSP ($\nu_{pk}^{syn} < 10^{14} Hz$), ISP ($10^{14} < \nu_{pk}^{syn} < 10^{15} Hz$) or HSP ($\nu_{pk}^{syn} < 10^{15} Hz$) blazars. On average, FSRQs are more energetic than BL Lacs. They are seen at higher redshifts, and present a more dramatic variability pattern \citep{Chiaro}, though some observational bias may be responsible for this trend. More specifically, FSRQs are surrounded by a relatively dense shell of gas and radiation (typically $\sim$~0.2~pc from the SMBH), which is called the broad-line region (BLR), due to the high velocity of these ``clouds'' that cause a Doppler broadening of the emission lines in the optical spectrum of the blazar. Consequently, it is easier to determine the redshifts of FSRQs, than those of BL Lacs whose spectra are often quasi featureless (their optical lines having an equivalent width $<$ 5 \AA{} in their rest frame). Most of the FSRQs belong to the LSP category, and rarely to the ISP category, though BL Lacs are fairly uniformly distributed within the three LSP, ISP and HSP categories \citep{2010ApJ...716...30A}.

The study of blazars is of particular importance for astrophysics and cosmology, as well as for particle and plasma physics. Physical processes involved in blazars under extreme conditions of acceleration, density, magnetic and gravitational fields can be probed only in such objects. Also, since the redshifts of gamma-ray blazars spread from $\sim$0.1 to $>$5, blazar populations constitute probes to study the evolution of galaxies throughout the history of the Universe. Improvement in the blazar classification is then of significant importance for addressing the physical, astrophysical and cosmological challenges mentioned above.

%%%%%%%%%%%%%%%%%%%%%%%%%%%%%%%%%%%%%%%%%%%%%%%%%%%%%%%%%%%%%%%%%%%%%%%%%%%%

The {\it Fermi} Gamma-Ray Space Telescope has orbited the Earth since June 2008. It operates in survey mode for most of the time, covering the whole sky every 3 h (corresponding to two orbits), thanks to its 2.4-sr field of view. This allows a regular monitoring of sources on the whole sky. However, due to a technical problem with the spacecraft, scientific operations were interrupted from 16 March till 8 April 2018. {\it Fermi} resumed its mission from this date in partial sky-scanning mode\footnote{\url{https://fermi.gsfc.nasa.gov/ssc/observations/types/post\_anomaly/}.}. Its main instrument, the Large Area Telescope (LAT), is sensitive to photons from $\sim$20~MeV to $> 300$~GeV \citep{Atwood}.

During the first ten years of observation (2008--2018), 5788 sources above 4~$\sigma$ significance were detected above 50 MeV and were reported in the Fourth {\it Fermi}-LAT point source catalogue -- Data Release 2 \citep[4FGL--DR2]{4FGL,4FGL-DR2}. Among these sources, 3492 were identified as AGNs and are recorded in the Fourth {\it Fermi}-LAT AGN Catalog -- Data Release 2 \citep[4LAC--DR2]{4LAC,4LAC-DR2}. Most of the 4LAC--DR2 objects are blazars (3420 sources--including both the high and low Galactic latitude samples): 1190 BL Lacs, 733 FSRQs and 1497 blazar candidates of uncertain type (BCUs). In the {\it Fermi}-LAT  Third AGN Catalog (3LAC), BCUs are defined as objects that were categorised as blazars by other means than a detailed optical spectrum, and so are not yet classified as a BL Lac or FSRQ \citep{3LAC}. BCUs are subdivided into three subclasses --- from I to III, ``BCU I'' being the class with more information on the source, ``BCU III'' the class with the least information:
\begin{enumerate}
\item BCUs I: A published optical spectrum of the object is available, but not sensitive enough for a classification as an FSRQ or a BL Lac. Redshifts are often known.
\item BCUs II: the object lacks an optical spectrum, but a reliable position of the SED synchrotron peak can be determined. Redshifts are not known.
\item BCUs III: the object is lacking both an optical spectrum and an estimated synchrotron-peak position, but it shows blazar-like broadband emission and a flat radio spectrum. Redshift are not known.
\end{enumerate}
\sourceName is a radio source from the NRAO VLA Sky Survey (NVSS) catalogue \citep{Condon}, and is also recorded as WISE J141922.55-083832.0 in the catalogue of the {\it Wide-field Infrared Survey Explorer} in mid-infrared bands \citep{WISE}. It was classified as a BCU II in the 3LAC and although not a bright {\it Fermi} source, it was reported in the Second {\it Fermi}-LAT point source catalogue \citep[2FGL]{2FGL}, from 2 years of observations, at a significance level of $\sim 7 \sigma$ (2FGL J1419.4-0835). In the Third {\it Fermi}-LAT point source catalogue \citep[3FGL]{3FGL}, from 4 years of observations, it has a significance level of $\sim 15 \sigma$ (3FGL J1419.5-0836). Finally in the 4FGL and 4FGL--DR2 catalogues, from 8 and 10 years of observation, respectively, \NVSS is labeled as 4FGL J1419.4-0838 and has a significance of $\sim 47$ and $\sim 54 \sigma$, respectively \citep{4FGL,4FGL-DR2}.

\begin{figure*}
  \includegraphics[width=18cm]{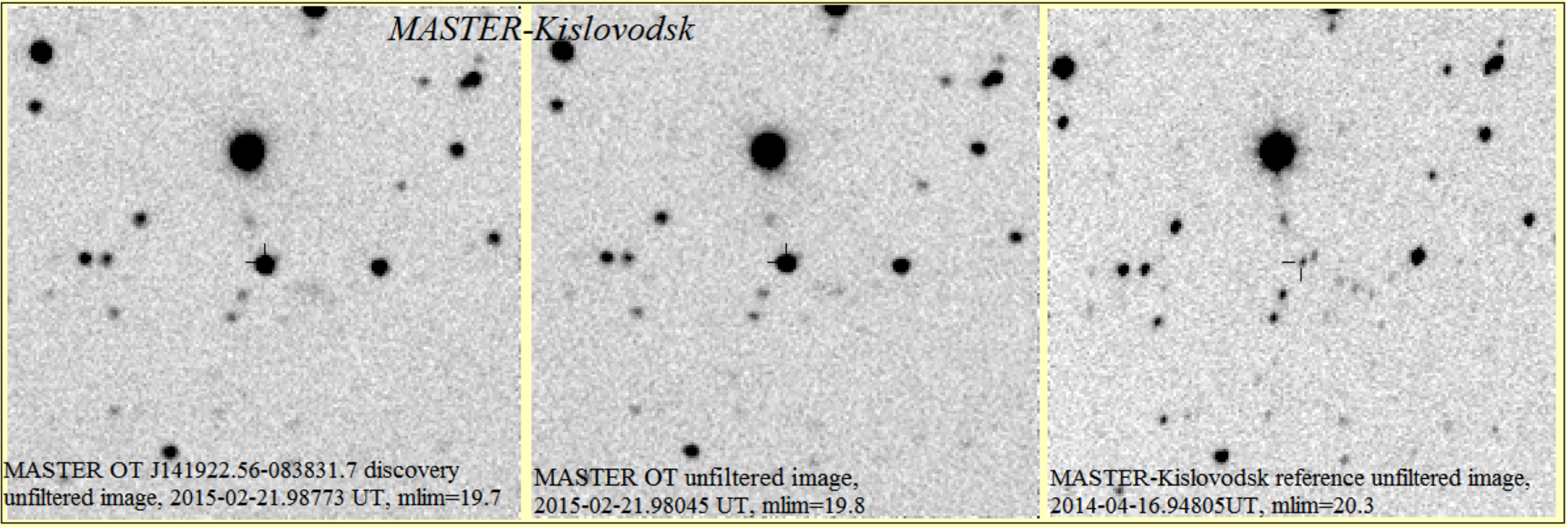} % updated 19 March 2022
  \caption{Images taken by MASTER-SAAO network. The first and second panels shows the unfiltered images observed on 21 February 2015 (MJD 57074) at different times, while the third is the reference image taken on 6 February 2014 (MJD 56694). The white-light (filtered) magnitude of NVSS J141922-083830 was measured at 14.6 on 21 February 2015 (MJD 57074.98773, first panel)}.
    \label{Figure-1:disc-image}
\end{figure*}

\NVSS was identified as an LSP source in the {\it Fermi}-LAT Second AGN Catalog \citep[2LAC]{2LAC}. In preparation for the 8-year 4LAC, the {\it Fermi}-LAT Collaboration performed new broad-band SED fits of many sources, which yielded for this source a synchrotron peak frequency $\nu_{syn} = 1.20 \times 10^{13}$ Hz at ($\nu F_\nu)_{syn} = 1.33 \times 10^{-12}$ erg cm$^{-2}$ s$^{-1}$ \citep{4LAC}. Several interesting studies were undertaken within the previous decade to identify blazar candidates as either BL Lacs or FSRQs from statistical tools using multiwavelength photometry \citep[e.g.][]{D'Abrusco,Chiaro,2018arXiv180805881C}. Though some results were achieved through these methods in identifying \NVSS as an FSRQ at some level of probablity, confirmation was still needed though a more complete study, including optical spectroscopy. We consider that this was achieved through our initial discovery alert \citep{2015ATel.7167....1B} with our first optical spectrum, plus the work presented in the current paper. Based on this initial work, the source type is referenced as a FSRQ in the 4FGL and subsequent catalogues.

In February--March 2015, a major flare of this source was discovered in the optical with the {\it MASTER Global Robotic Net} \citep{Lipunov2015}, in the near infrared with the {\it Guillermo Haro Observatory} \citep{Carrasco} and in X rays with the {\it Niel Gehrels Swift Observatory} \citep{2015ATel.7184....1C}. Following the MASTER alert, we triggered a spectroscopic observation with the Southern African Large Telescope (SALT). Our analysis of the {\it Fermi}-LAT data during this epoch allowed us to observe contemporaneous gamma-ray activity, along with the {\it Swift} observations. 
The broadband emission during this major flare is studied within the framework of theoretical emission models.\\
%Though we studied other flares in this Paper, we focused more on the February--March 2015 flare.}\\

The outline of the paper is as follows: in section 2 we present all of the optical observations which includes photometry, polarimetry and spectroscopy; in Section 3 we discuss the {\it Swift} XRT and UVOT observations; in Section 4 we present the {\it Fermi}-LAT observations and analysis; in Section 5 the SED and spectral modelling results are presented, while finally the summary and conclusions are given in Section 6.

%%%%%%%%%%%%%%%%%%%%%%%%%%%%%%%%%%%%%%%%%%%%%%%%%%%%%%%%%%%%%%%%%%%%%%%%%%%%
\section{Optical Observations and Data Analysis }

Here we report on the optical observations of \NVSS. These extend over a period of 7.7 years and cover three periods when \NVSS underwent flaring episodes. The optical observations include photometry and polarimetry from three MASTER observatories, in South Africa (MASTER-SAAO), Russia (MASTER-Kislovodsk) and the Canary Islands (MASTER-IAC), plus spectroscopy from SALT.

\subsection{MASTER optical observations} \label{sec:MASTER}

 The MASTER network \citep{Lipunov2010} consists of 9 identical fully robotic telescope systems situated at different nodes, five in Russia (MASTER-Amur, -Tunka, -Ural, -Kislovodsk, -Tavrida), MASTER-SAAO at the SAAO, MASTER-IAC in Spain at Instituto de Astrof\'isica de Canarias (IAC, Tenerife), MASTER-OAFA at the Observatorio Astronomico Felix Aguilar of San Juan National University in Argentina and MASTER-OAGH in Mexico.  Each node consists of identical dual 0.4-m diameter wide field telescopes (Hamilton design) on a common mount, with Apogee CCD cameras (4k $\times$ 4k $\times$ 9$\mu$m pixels) with BVRI and clear filters and 2 broad-band polarizers \citep{Kornilov2012}. The telescope tubes can either be co-aligned, each surveying an identical $2^{\circ} \times 2^{\circ}$ field, with two different filters or polarizers, or they can be misaligned to allow twice the sky coverage using identical filters. The latter is the default for the transient survey mode, which is usually done without using a filter to maximize sensitivity \citep{Lipunov2010,Lipunov2019a,Lipunov2022,Kornilov2012}. 
 
Each MASTER observatory is controlled by identical automated software (the main MASTER key feature) allowing them to operate as both autonomous fast alert follow-up systems and as survey telescopes for the discovery of astrophysical transients of all classes. 
Data reduction tasks, which include standard dark, flat field and astrometry calibrations, are performed using the MASTER pipeline software, where dark frames and twilight flats are  obtained in automatic mode. Aperture photometry of the object and comparison stars was performed using Photutils \citep{Bradley}. 
 
The MASTER software gives information about all optical sources from images 1-2 minutes following CCD readout. This information includes the full classification of all sources from the image, the full dataset from previous MASTER-Net archive images for every sources, the full information from VIZIER database and all open sources (e.g. Minor Planet Center), orbital elements for moving objects, etc. Taking into account the large field of view of the MASTER telescopes 2 x (2$^{\circ}$ x 2$^{\circ}$), we have at our disposal a large number ($\approx$ 3000 to 10000) of reference stars to use for photometric reductions. As a result, the photometric errors are minimized. Comparison stars for automatic initial photometry are typically chosen from the USNO-B1 \citep{USNO_B} catalogue and the USNO magnitudes transform to the MASTER unfiltered system as W=0.2B + 0.8R . Errors are determined from the fitting relations between the standard deviations and magnitudes for the comparison stars\citep{Lipunov2010,Kornilov2012,Gorbovskoy}.
 
 Each pair of MASTER telescope tubes are equipped with achromatic wire-grid polarizers, mutually perpendicular to one another. In the case of the two MASTER observatories which obtained contemporaneous polarimetry of \NVSS (3$-$5 March 2015), for MASTER-Kislovodsk, the planes are at positional angles 0$^{\circ}$ and 90$^{\circ}$ to the celestial equator (this is also the same for MASTER-IAC), while for MASTER-SAAO, the angles are 45$^{\circ}$ and 135$^{\circ}$. 
 Using both contemporaneous MASTER observations, with the four different orientations of the polarizers, it is possible to derive the $Q$ and $U$ Stokes parameters and hence the degree of linear polarization and position angle. In spite of the fact that all MASTER telescopes have a similar construction, optical scheme, and polarizers, there are some differences in the CCD responses which limit the precision of the polarization measurements. Also, the calibration using known polarized galactic sources is not possible, since any measurement of polarization involves the subtraction of nearby field stars between at least two images, obtained with tubes with differently oriented polarizers. Thus, the derived polarization of a target object is always relative to the nearby field stars. They are usually chosen from the area around the target source in a radius of 10 arcmin and are bound to have the same polarization due to the similar interstellar matter properties toward them \citep{Kornilov2012,Gorbovskoy2016,Lipunov2019b}.
 
 \begin{figure}
  \begin{center}
    \includegraphics[trim=0.9cm 0.1cm 1.2cm 1.2cm, clip, width=\columnwidth]{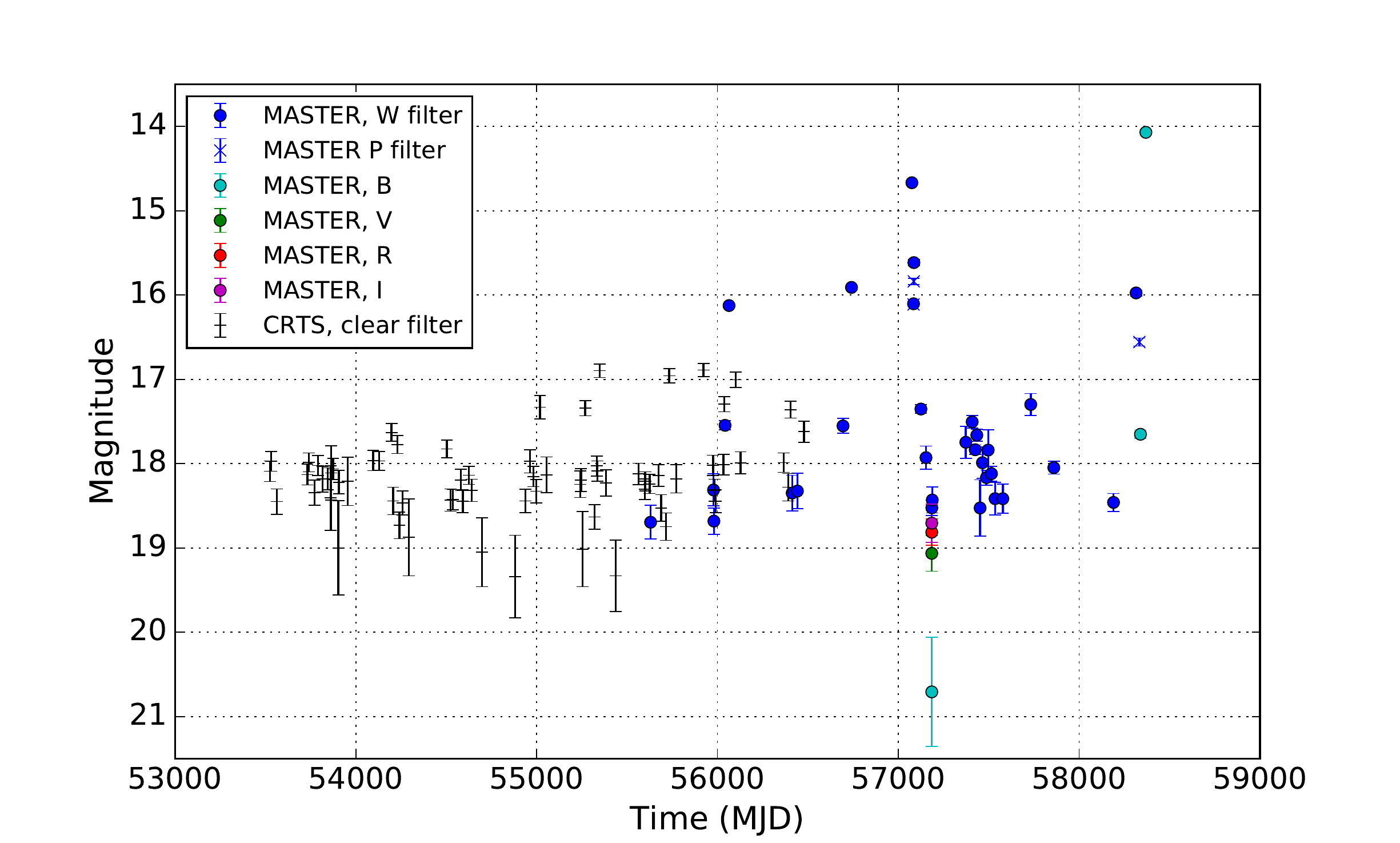}
  \caption{MASTER network light-curves of NVSS J141922-083830, together with measurements from CRTS, all binned in one day intervals.\label{Figure-2:longterm-phot}}%\label{Figure-2:longterm-phot}
  \end{center}
\end{figure}
 
\cite{Lipunov2015} reported the detection of an optical flare from \sourceName using the MASTER optical transient auto-detection system on 2015-02-21.98773 UTC (MJD 57074.988) as MASTER OT J141922.56-083831.7. The unfiltered flare magnitude was 14.6 and the MASTER-Kislovodsk discovery images plus reference image are shown in Figure \ref{Figure-1:disc-image}, The source was seen in 60 previous MASTER images, obtained since 9 March 2011 (MJD 55629). The quiescent brightness values, over 3 years (2011-2014), range from 17$-$19. 
A total of three additional flares have been detected in MASTER archive images. The unfiltered magnitudes of these flares were 16.1, 15.9 and 14.5$-$16.5, respectively.
Flares from \NVSS were first reported from the Catalina Real Time Survey \citep[CRTS;][]{CRTS}, where it was observed to reach a maximum brightness of V = 16.85 around MJD 55017 \citep[CSS100601:141922-083830;][]{Djorgovski}, which are shown in Figure \ref{Figure-2:longterm-phot}.

\subsubsection{Followup optical photometry}

Followup photometric observations of \sourceName were obtained on MASTER-SAAO and MASTER-Kislovodsk, starting $\approx$9 d after the flare detection (from 2 March 2015/MJD 57083.964, when it appears the {\it Fermi}-LAT flux had dropped by a factor of $\approx$2). 
Simultaneous observations on two optical tubes in the clear $W$ filters were processed as independent observations, while fluxes u sing the two orthogonal polarizers of one MASTER telescope pair were summed as one observation. After this, photometric data processing was performed with Astrokit \citep{Burdanov} to minimize the standard deviation of an ensemble of comparison stars chosen from the Pan-STARRS1 (PS1) catalogue \cite{Chambers,Kostov}.

\begin{figure}
   \begin{center}
     \includegraphics[trim=0cm 0cm 0cm 0cm, clip, width=\columnwidth]{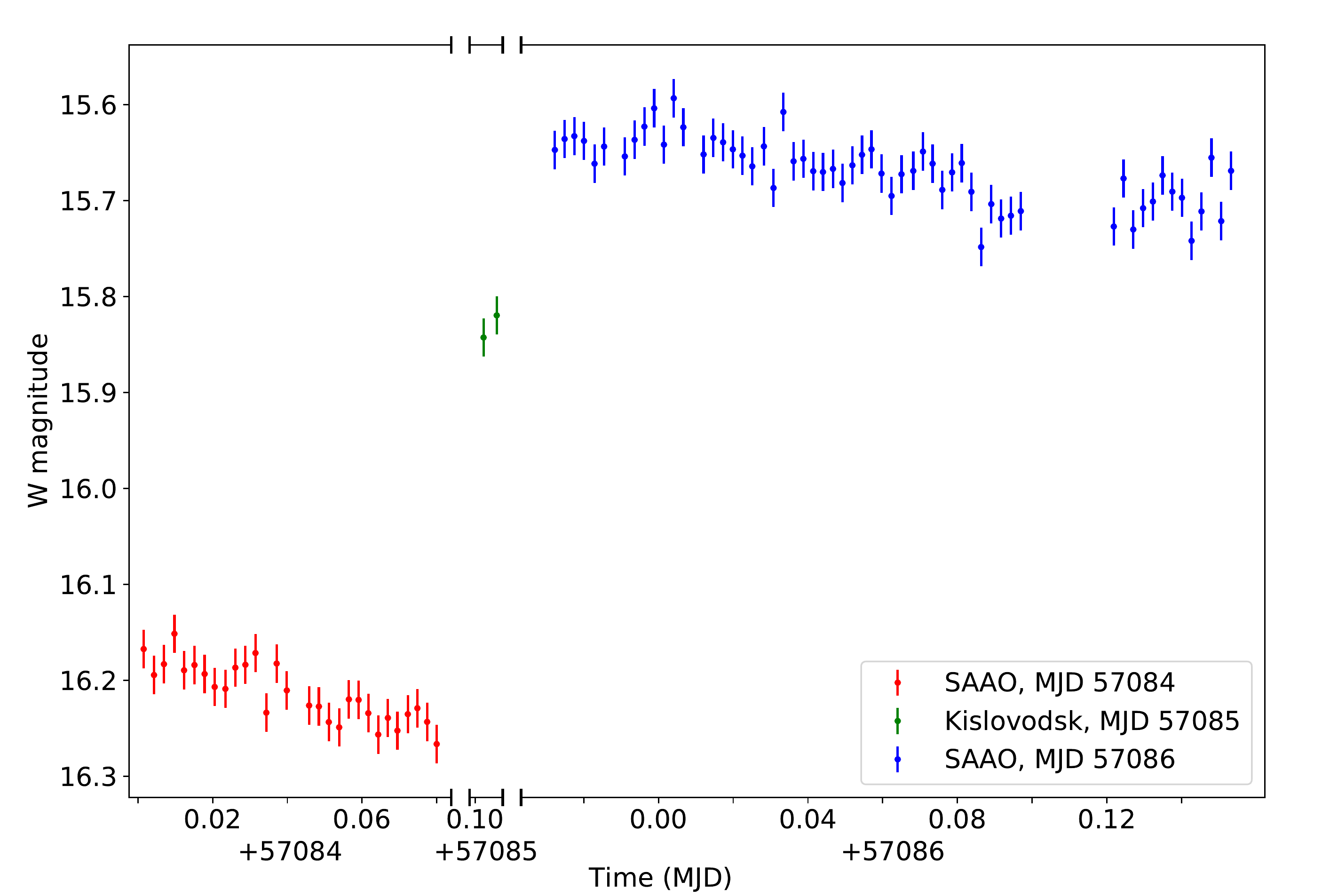}
   \caption{High time resolution light curve based on observations with polarization filters at 3--5 March 2015 (MJD 57084--57086), just past the peak of gamma-ray Flare~3 (see Section \ref{Fermi}). 
   %Note that each comparison stars has three measurements. 
   }\label{Figure-3:Mar2015-phot}
   \end{center}
 \end{figure}

The day-binned light curve of \sourceName is presented in Figure \ref{Figure-2:longterm-phot}. Errors were determined as the standard deviation of magnitudes in each bin. The longest continuous sets of photometric observations were obtained on 3--4 and 5 March 2015 (MJD 57084--57086). The brightness of the source was seen to slowly decrease during the SAAO 3--4 March observation, while observations on 4 \& 5 March showed the source to have brightened by $\approx$ 0.5 magnitude. The light curve over this period is shown in Figure~\ref{Figure-3:Mar2015-phot}.

\sourceName was observed briefly on 1 March 2015 (MJD 57082.00) using the SAAO 1.9 m telescope and the SHOC high-speed EM-CCD camera. Measurements confirmed the object was still in a bright flaring state (V $<$ 15), the motivation for a subsequent SALT spectroscopic observation (see Section \ref{spec}).

\begin{figure}
  \begin{center}
    \includegraphics[trim=0.3cm 0.2cm -0.5cm 1.2cm, clip, width=1.1\columnwidth]{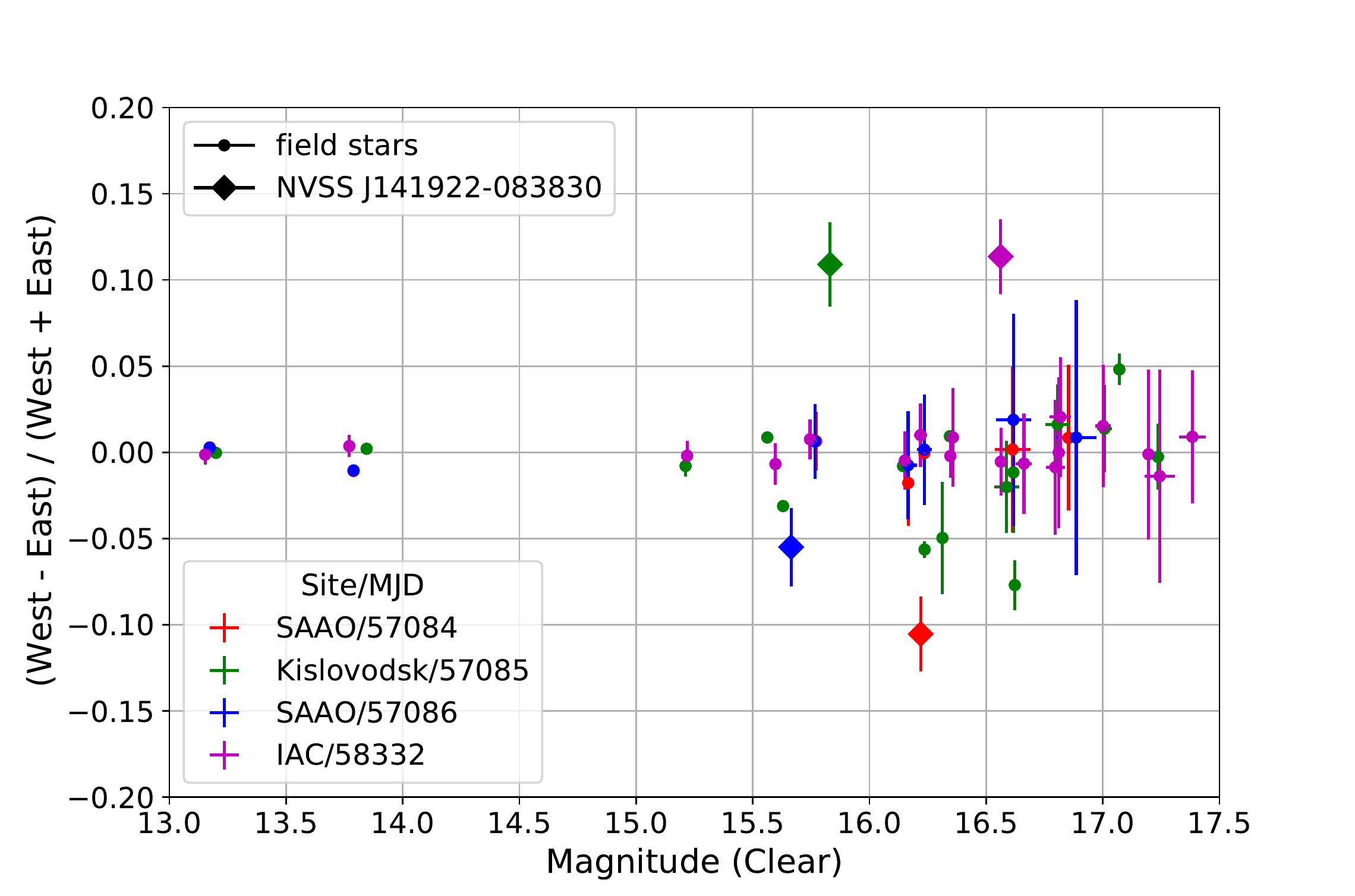}
  \caption{
  MASTER measurements of normalized Stokes $q$ and $u$ parameters for NVSS J141922-083830 during two flaring episodes, in March 2015 (SAAO (u) and Kislovodsk (q)) and August 2018 (IAC (q)), together with similar measurements for foreground stars in the field. These are determined by taking the ratio of the difference and summed intensities, respectively, observed in each pair of CCD frames corresponding to the orthogonally offset polarizers.
  \label{Figure-4:polarization}
  }
  \end{center}
\end{figure}

\subsubsection{Polarization measurements}
Polarimetric observations were carried out in March 2015 and August 2018 with the MASTER-SAAO, MASTER-Kislovodsk and MASTER-IAC observatories. Details are presented in Table \ref{tab-pol}. We see that \NVSS is consistently polarized, from all four attempted measurements. 

The degree of polarization, $P$ was calculated as $(\mathrm{ Flux_{West} - Flux_{East}) / (Flux_{West} + Flux_{East}})$. The estimated values correspond to normalized Stokes parameters q for Kislovodsk and IAC observations and u for SAAO observations. The results are presented in Figure \ref{Figure-4:polarization} for both \NVSS and a number of field stars.  As can be seen from the plot, only three field stars showed any evidence for being polarized, not surprisingly given the high Galactic latitude ($b$ = 47.83$^{\circ}$). We also investigated any potential correlations between the measured minimum polarization from the MASTER-SAAO and MASTER-IAC data with Gaia DR2 parallaxes and found none.

\begin{figure}
  \begin{center}
    \includegraphics[trim=0.9cm 0.1cm 1.2cm 1.2cm, clip, width=\columnwidth]{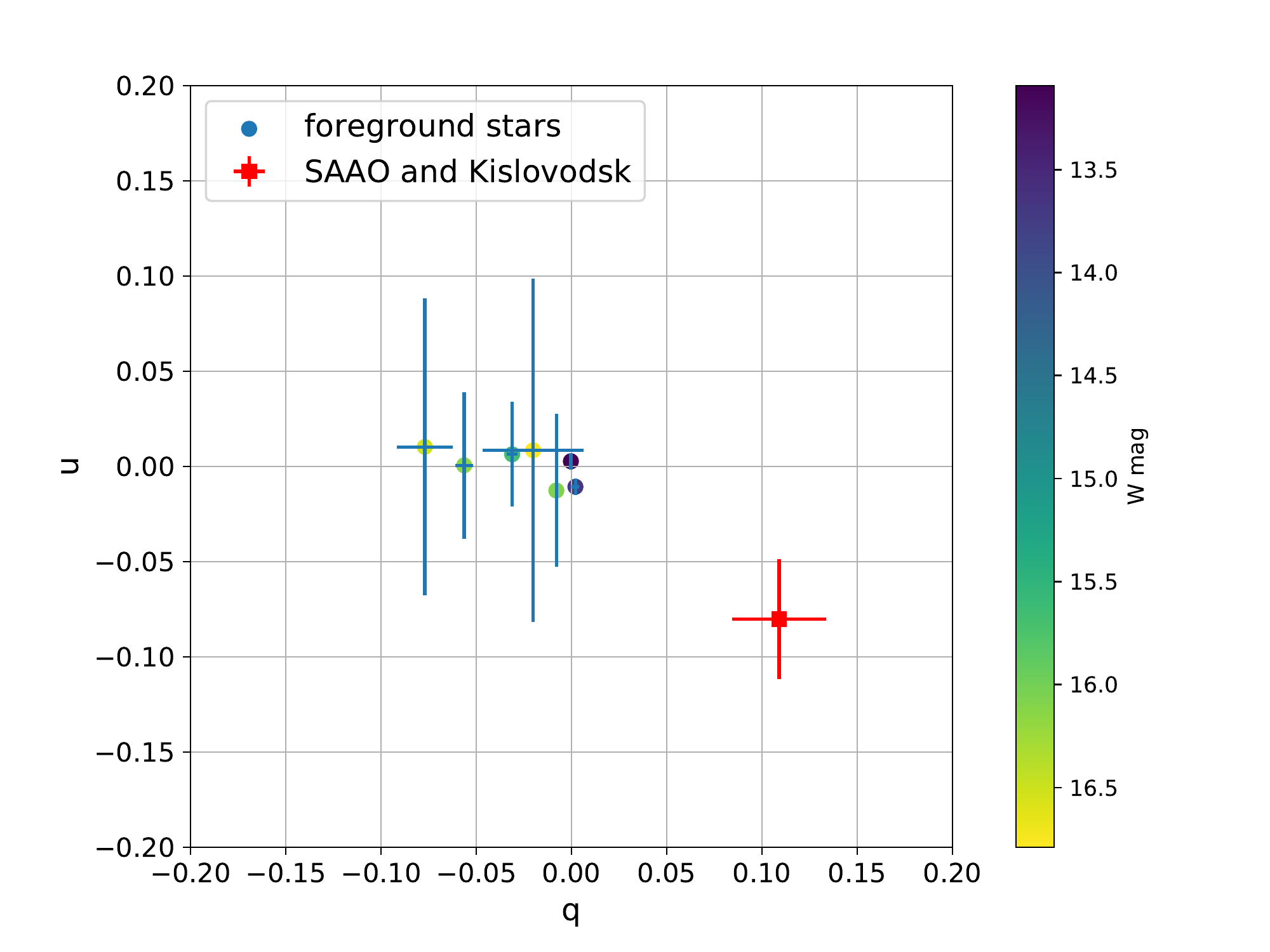}
  \caption{Normalized Stokes $q$ and $u$ parameters determined for \NVSS and nearby field stars during the flare episodes in March 2015 (gamma-ray Flare 3 -- see Section \ref{Fermi}).}\label{Figure-5:qu}
  \end{center}
\end{figure}

\begin{table*}
  \begin{center}
  \caption{Log of polarization observations of \NVSS}  \label{tab-pol}
  \begin{tabular}{lccccc}
    \hline
    Observatory     & Date Range    & MJD range & \# of CCD exposure pairs & $|P|$ & W magnitude \\
    \hline
    MASTER-SAAO &	2015-03-03.002 to 2015-03-03.080 & 57084.002--57084.080 & 29 & 10.5 $\pm$ 2.2\% & 16.22 $\pm$ 0.03\\
    MASTER-Kislovodsk & 2015-03-04.102 to 2015-03-04.104 & 57085.102--57085.104 &  2 & 10.9 $\pm$ 2.5\% & 15.83 $\pm$ 0.01\\
    MASTER-SAAO &	2015-03-04.972 to 2015-03-05.153 & 57085.972--57086.153 & 59 & 5.5 $\pm$ 2.3\% & 15.66 $\pm$ 0.03\\
    MASTER-IAC &	2018-08-02.880 to 2018-08-02.906 & 58332.880--58332.906 & 10 & 11.4 $\pm$ 2.2\% & 16.56 $\pm$ 0.04\\
    \hline
  \end{tabular}
  \end{center}
\end{table*}

In the absence of polarimetric observations from two or more MASTER observatories overlapping in time, we can only determine a lower limit for the degree of polarization. Hovever, if we assume a linear change of the degree of polarization, $p$, between the two MASTER-SAAO observations on 3 \& 5 March 2015, plus a constant polarization angle, $\theta$, then we derive the mean normalized Stokes parameters for the time of the Kislovodsk polarimetric observation, performed in between the two MASTER-SAAO observations.
The combined observations were obtained with 4 angles of the polarizers, namely 0, 45, 90 \& 135$^{\circ}$, which allows us to determine the normalized Stokes $q$ and $u$ parameters for \NVSS and 7 field stars, which are shown in Figure \ref{Figure-5:qu}. We determined these to be $q  = 0.109 \pm 0.025$ and $u = -0.080 \pm 0.031$, implying $p = (13.5 \pm 3.9)\%$ and $\theta = (-18.2 \pm 12.5) ^{\circ}$.

%%%%%%%%%%%%%%%%%%%%%%%%%%%%%%%%%%%%%%%%%%%%%%%%%%%%%%%%%%%%%%%
\subsection{SALT Optical spectroscopy}  \label{sec:spectro}
%%%%%%%%%%%%%%%%%%%%%%%%%%%%%%%%%%%%%%%%%%%%%%%%%%%%%%%%%%%%%%%
\label{spec}
An optical spectrum of \NVSS was obtained with SALT \citep{Buckley2006} on 1~March 2015 \citep[MJD 57082;][]{2015ATel.7167....1B}, with the Robert Stobie Spectrograph \citep[RSS;][]{Burgh2003}. The PG900 VPH grating was used at a grating angle of 14.0$^{\circ}$. A 1100~s exposure spectrum, covering 3780--6850\AA{} at a resolution
of 4.8\AA{} with a 1.25 arcsec slit, was obtained in clear conditions and seeing of 1.5 arcsec. 

The data were reduced using the PySALT package, a PyRAF-based software package for SALT data reductions \citep{Crawford2010}\footnote{\url{https://astronomers.salt.ac.za/software/pysalt-documentation/}}, which includes corrections for both gain and cross-talk, bias subtraction, amplifier mosaicing, and removal of cosmetic defects. The individual spectra were then extracted using standard
IRAF\footnote{\url{https://iraf.noao.edu/}} procedures, wavelength calibration (with a calibration lamp exposure taken immediately after the science spectra), background subtraction and extraction of 1D spectra. We could only obtain relative flux calibrations, from observing spectrophotometric standards in twilight, due to the SALT design \citep{Buckley2018}.  

%%%%%%%%%%%%%%%%%%%%%%%%%%%%%%%%%%%%%%%%%%%%%%%%%%%%%%%%%%%%%%%
\begin{figure} %%% FIGURE 6
\begin{center}
  \includegraphics[width=8.5cm]{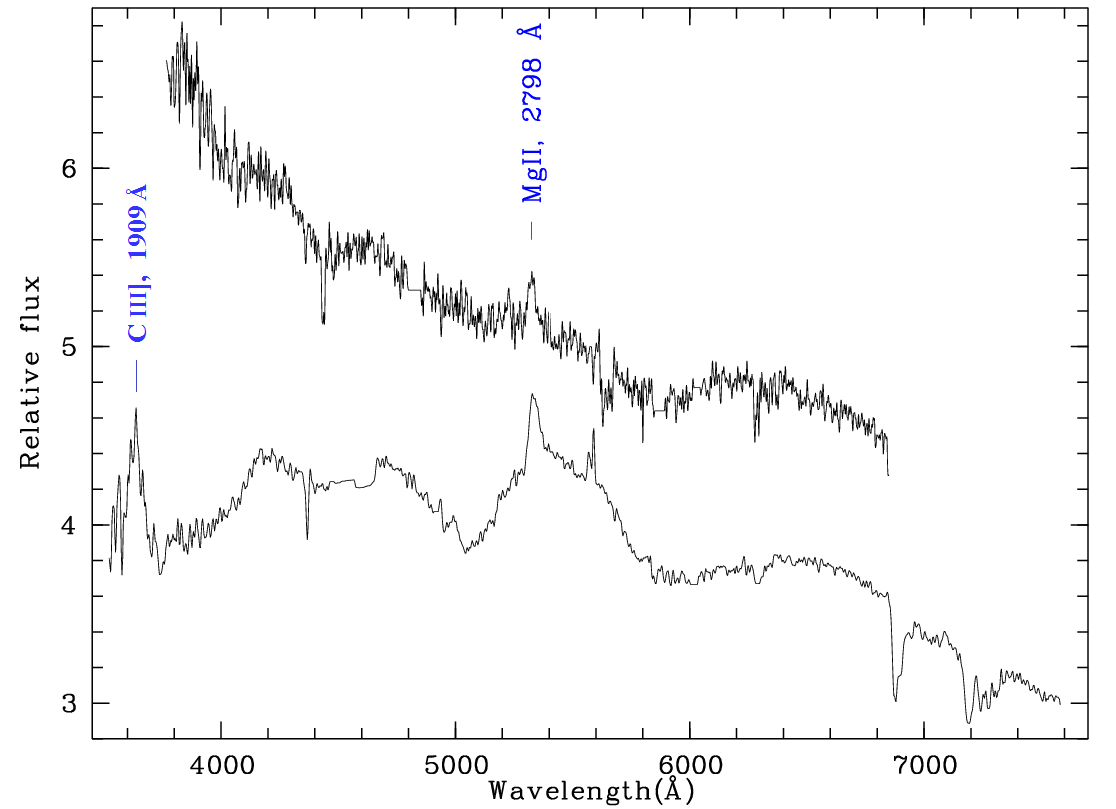}
\end{center}
  \caption{\label{spec1} SALT spectra of NVSS J141922-083830 obtained on 1 March 2015 (MJD 57082; upper), 7 days after the flare detection, and 30 May 2017 (MJD 57903; lower), when in quiescence. The emission feature at 5325\AA~is identified as Mg~{\sc ii} 2798\AA, which implies a redshift of z=0.903. The line towards the blue end of the 2017 observation is identified as the C~{\sc iii}] 1909\AA{} line. Note that the flux levels are relative and the 2017 spectrum (lower plot) has been scaled by a factor of 1.5.}\label{Figure-6:spectra}
\end{figure}
%%%%%%%%%%%%%%%%%%%%%%%%%%%%%%%%%%%%%%%%%%%%%%%%%%%%%%%%%%%%%%%

The spectrum, included in Figure \ref{Figure-6:spectra}, shows a single emission line at
5325\AA\/ with an equivalent width of E.W. = $1.25 \pm 0.11$ and line width of FWHM = $28.1 \pm 1.8$ \AA{}. The only other spectral features worth commenting on are some very broad ($\sim$500\AA{}) bumps at $\sim$4200, 4600 and 6400\AA, superimposed on a continuum which steeply rises to the blue. Some narrow absorption features are seen, which  are due to detector artifacts from imperfect bad pixel masking.

If the emission line is interpreted to be the Mg~{\sc ii} 2798\AA{} line, then the implied redshift for this blazar is 0.903. Other possible lines, which are  typically seen in blazars (e.g. Ly$\alpha$1216\AA, O~{\sc iv}]/Si~{\sc iv}1400\AA{}, C~{\sc iv}1549\AA{} and C~{\sc iii}]1909\AA) are ruled out, since we would expect to see more than one emission line in the our spectrum for the corresponding redshifts.

Further SALT RSS spectra of \sourceName were obtained on 30 May 2017 (MJD 57903) when the source was quiescent in terms of $\gamma$-ray emission. Comparison of the SALTICAM acquisition images taken during the two epochs of the SALT observations (2015 \& 2017) indicates that the source was $\sim$2 magnitudes fainter for the latter observation, i.e. optically fainter by a factor of 6.3. Two spectra of 1200 s exposures each were obtained, in clear conditions with median seeing of 1.6 arcsec, using the lowest resolution PG300 surface relief transmission grating (300 lines/mm), which covered 3200--8900\AA{} at a mean resolution of 17\AA{} with a 1.5 arcsec slit. This spectrum (Figure \ref{Figure-6:spectra}) also shows the same Mg~{\sc ii} 2798\AA{} line seen in 2015, with an E.W. = 4.5\AA. Additionally another emission line is seen at the blue end, at $\approx$3630\AA{}. We identify this as the C~{\sc iii}] 1909\AA{} line, which is consistent with a redshift of $z$ = 0.903. This is further evidence that the observed emission line in the first spectrum was Mg~{\sc ii} 2798\AA{}, as previously reported. The broad continuum humps previously seen in the 2015 spectrum are also seen in the 2017, and are even more pronounced. The SALT spectra shown in Figure \ref{Figure-6:spectra} show relative fluxes, with the 2017 spectrum scaled by a factor of 1.5.

%%%%%%%%%%%%%%%%%%%%%%%%%%%%%%%%%%%%%%%%%%%%%%%%%%%%%%%%%%%%%%%%%%%%%%%%%
\subsection{{\it Neil Gehrels Swift Observatory}-UVOT observations}
%%%%%%%%%%%%%%%%%%%%%%%%%%%%%%%%%%%%%%%%%%%%%%%%%%%%%%%%%%%%%%%%%%%%%%%%%

The UVOT observations are analysed using the recent mission specific tools {\tt uvotimsum}, {\tt uvotsource} and {\tt uvot2pha} distributed with the {\it heasoft} package. The sky images in a particular filter corresponding to individual observations are combined using {\tt uvotimsum} to get a single frame per observation, whenever more than one image was taken. The combined images are then analysed utilizing the tool {\tt uvotsource} using a circular region of 5" radius centred at the sky location of \NVSS as the source region. Another circular region of 35.76" located in a source free region around 3.5' away from \NVSS is used to extract background counts. 

A correction due to reddening, E(B-V)=0.178, due to the presence of the neutral hydrogen along the line of sight within our own Milky Way Galaxy, is applied to the fluxes before using these values for the SEDs. The reddening is estimated by the Python module {\tt extinctions} using the two-dimensional dust map of the entire sky by \citet{1998ApJ...500..525S} which was recently updated by \cite{2011ApJ...737..103S} [SFD hereafter]. The estimation of the same parameter using the two-dimensional dust map at NASA/IPAC archive\footnote{\url{https://irsa.ipac.caltech.edu/applications/DUST/}} yields a value of 0.038. The empirical formalism by \cite{1989ApJ...345..245C} 
with A$_V$ = R$_V$ * E(B-V) and R$_V$ = 3.1 is used to estimate the correction factor A$_\lambda$ for individual UVOT filters. 

%%%%%%%%%%%%%%%%%%%%%%%%%%%%%%%%%%%%%%%%%%%%%%%%%%%%%%%%%%%%%%%%%%%%%%

%%%%%%%%%%%%%%%%%%%%%%%%%%%%%%%%%%%%%%%%%%%%%%%%%%%%%%%%%%%%%%%%%%%%%%
\section{{\it Neil Gehrels Swift Observatory} X-ray Telescope (XRT)\label{subsec:swiftobs}}
 
 \begin{table}
     \centering
     \caption{{\it Swift} observations of \NVSS.\label{swobs}}
     {\footnotesize \begin{tabular}{llll}\hline
      ObsID & Date-OBS & Exposure & UVOT
        filters\\
            &  [UTC]   &  (s)     &      \\
        \hline
     00046513001  & 2013-06-24 & 756.7 & W2, M2, W1, U, B, V\\
     00046513003  & 2014-09-07 & 359.6 & W1 \\
     00046513004  & 2014-09-09 & 1171.2 & W2, M2, W1, U, B, V \\
     00046513005  & 2014-09-10 & 894.0  & W2, M2, W1, U, B, V\\ 
     00046513006 &  2014-09-10 & 299.7 & - \\ 
     00046513007 &  2015-03-04 & 1952.9 & W2\\\hline
     \end{tabular}}
     \label{tab:my_label}
 \end{table}
%%%%%%%%%%%%%%%%%%%%%%%%%%%%%%%%%%%%%%%%%%%%%%%%%%%%%%%%%%%%%%%%%%%%%%%%%%%%
\begin{table*}
  \begin{center}
  \caption{Time ranges of the five {\it Fermi}-LAT observation intervals. 
  }
  \label{tab:five_phases}
  \begin{tabular}{cccc}
    \hline
    Phase     & Date           & MJD            & MET\\
    \hline
    Quiescent & 4 Aug 2008--18 Apr 2013 & 54682.7--56400.0 & 239557417--387936003\\
    Flare 1   &	7 Feb 2014--2 Apr 2014 & 56695.0--56749.0 & 413424003--418089603\\
    Flare 2   &	13 Oct 2014--5 Nov 2014 & 56943.0--56966.0 & 434851203--436838403\\
    %Flare 3   &	14 Feb 2015--25 Feb 2015 & 57067.0--57078.0 & 445564803--446515203\\
    Flare 3   &	14 Feb 2015-- 9 Mar 2015 & 57067.0--57090.0 & 445564803-447552003\\
    Flare 4   & 31 Aug 2018--12 Sep 2018 & 58361.0--58373.0 & 557366405--558403205\\
    \hline
  \end{tabular}
  \end{center}
\end{table*}
%%%%%%%%%%%%%%%%%%%%%%%%%%%%%%%%%%%%%%%%%%%%%%%%%%%%%%%%%%%%%%%%%%%%%%%%%
\begin{figure*} %%% FIGURE 7
  \begin{center}
    %\label{Figure-7} 
    %\includegraphics[width=18cm]{Plot_LC_NVSS_J1419_New_3day_MJD_56599_58450_opt1_2021_input_set_141922_comp_0_Ind_UL_LCR_0_19March2022_OK.pdf} % updated 19 March 2022
    \includegraphics[width=18cm]{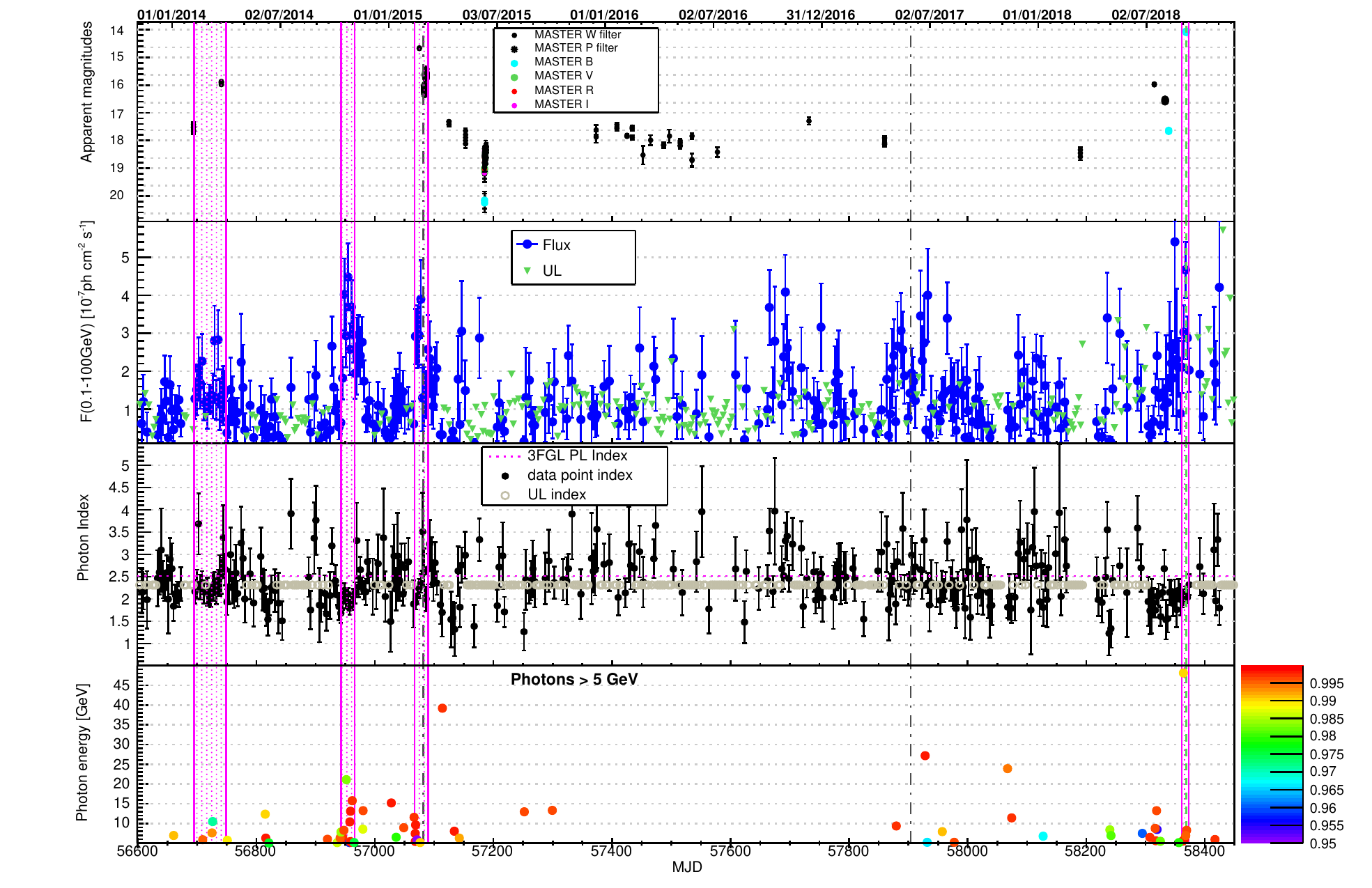} % updated 18 July 2022
  \caption{Light-curves, photon index and high-energy photons of NVSS J141922-083830 from 3 November 2013 to 28 November 2018 (MJD 56599 to 58450), encompassing the four flaring periods (highlighted with magenta shaded areas). The time stamp of the two SALT spectra are indicated by vertical grey dashed-dotted lines, and the time stamp of the MASTER-Net higher detected flux (7 September 2018 / MJD 58368.8545 at B=14.04) is indicated by a vertical green dashed line.
  Upper panel: MASTER-Net magnitudes from Figure 2. 
  Second panel: {\it Fermi}-LAT light-curve in the 100~MeV--100~GeV energy range, for a three-day binning. Upper limits are shown in green. 
  Third panel: power law photon index as optimised by the likelihood analysis for each bin of the {\it Fermi}-LAT light-curve. In light grey are indicated the values used for computing the upper limits (average values). Bottom panel: highest energy "{\sc ultraclean}" photons candidates $>$ 5 GeV. The colour code labeled from 0.95 to 1.00 on the right indicates the probability for each event to be associated to \NVSS.\label{Figure-7}} 
  \end{center}
\end{figure*}

%%%%%%%%%%%%%%%%%%%%%%%%%%%%%%%%%%%%%%%%%%%%%%%%%%%%%%%%%%%%%%%%%%%%%%%%%
\begin{figure*} %%% FIGURE 8
  \begin{center}
    \includegraphics[width=18cm]{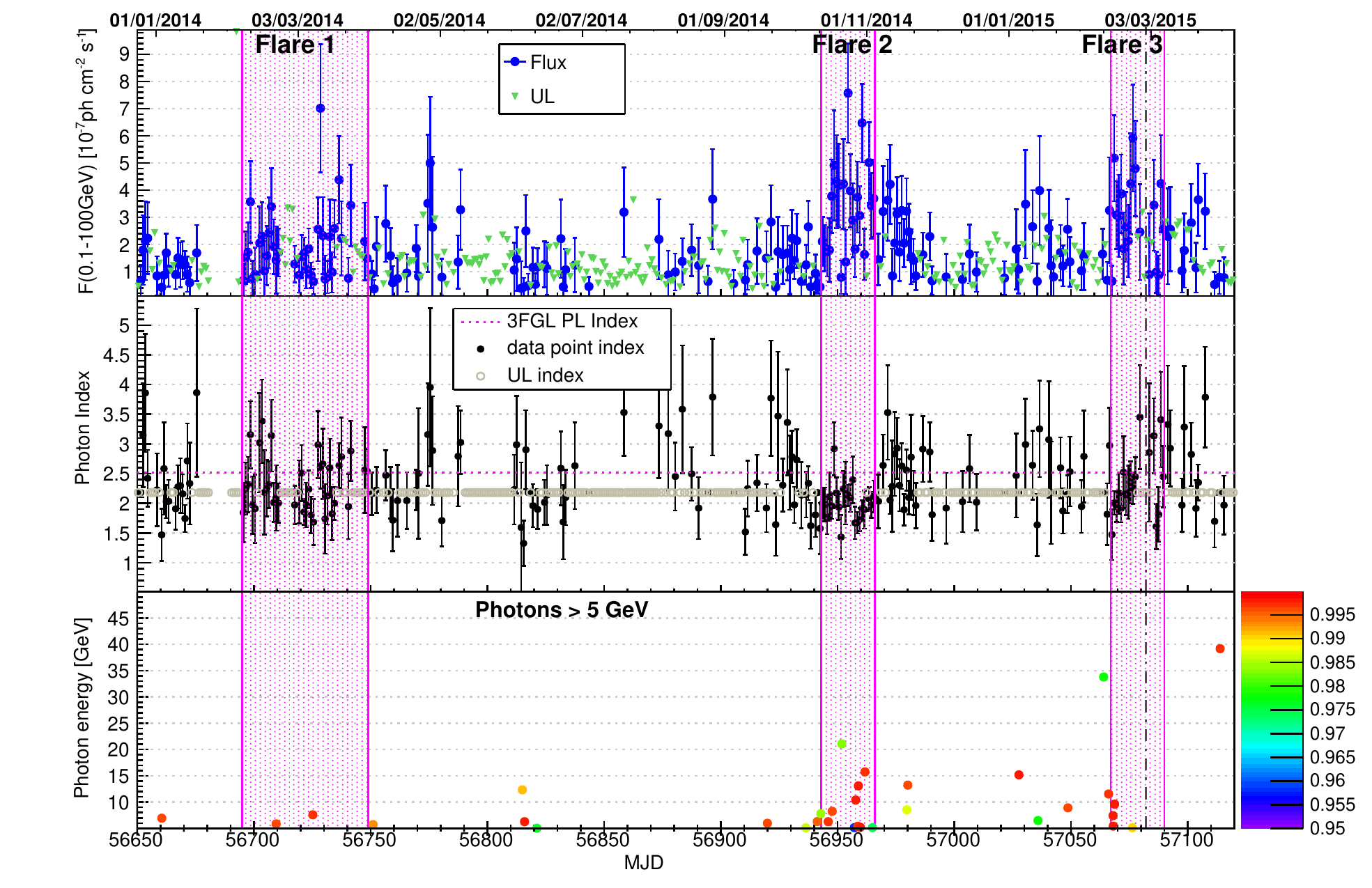} % updated 18 July 2022
  \end{center}
  \caption{\label{Figure-8} {\it Fermi}-LAT light-curve, photon index and high-energy photons of NVSS J141922-083830 from 24 December 2013 to 8 April 2015 (MJD 56650 to 57120), encompassing the first three flaring periods (highlighted with magenta shaded areas). The time stamp of the first SALT spectrum ($1^{st}$ March 2015) is indicated by a vertical grey dashed-dotted line. Upper panel: {\it Fermi}-LAT light-curve in the 100~MeV--100~GeV energy range, for a one-day binning. Upper limits are represented in green. 
  Second panel: power law photon index as optimised by the likelihood analysis for each bin of the {\it Fermi}-LAT light-curve. In light grey are indicated the values used for computing the upper limits (average values). Bottom panel: highest energy "{\sc ultraclean}" photons candidates $>$ 5 GeV. The colour code labeled from 0.95 to 1.00 on the right indicates the probability for each event to be associated to \NVSS.}
\end{figure*}
%%%%%%%%%%%%%%%%%%%%%%%%%%%%%%%%%%%%%%%%%%%%%%%%%%%%%%%%%%%%%%%%%%%%%%%%%%%%%%%%%%%%%%%%%%%%%%%%%%%%%%%%%%%%%%%%%%%%%%%%%%%%%%%%%%%%%%%%%%%%%%%%%%%%%
\swift\ performed six observations covering the interval 24 June 2013 (MJD 56468) to 4 March 2015 (MJD 57086), with details presented in Table \ref{swobs}. The X-ray data from XRT were reprocessed with the mission-specific {\it heasoft} tool {\tt xrtpipeline} (version 0.13.4) with standard input parameters as recommended by the instrument team. This step generates new cleaned events files with the most recent calibrations. The events files thus generated are used in the multi-mission tool {\tt XSELECT} for extracting source and background products. Four \swift\ observations of the \NVSS were performed in PC mode and two in WT mode. The source count rate was always $<$ 0.5 counts/s which confirms that the source is not affected by the pileup effect. A circular region of radius 75$^"$, centred at ($\alpha$=14:19:22.560, $\delta$=-08:38:32.20) is used as source region. Whereas, for the background, two circular regions each of radius 150$^"$ centered at ($\alpha$=14:19:27.57, $\delta$=-08:30:29.08) and ($\alpha$=14:19:57.636, $\delta$=-8:39:23.83) are used. We cross-checked for the presence of another X-ray source contaminating the source or background regions. The source being very faint, the two WT mode observations are not usable. Only the PC mode observations from OBSID 00046513007 are worth fitting the spectra and hence using for the X-ray part of the SED.

%%%%%%%%%%%%%%%%%%%%%%%%%%%%%%%%%%%%%%%%%%%%%%%%%%%%%%%%%%%%%%%%%%%%%%%
\begin{figure*}
\begin{center}
\includegraphics[width=15cm]{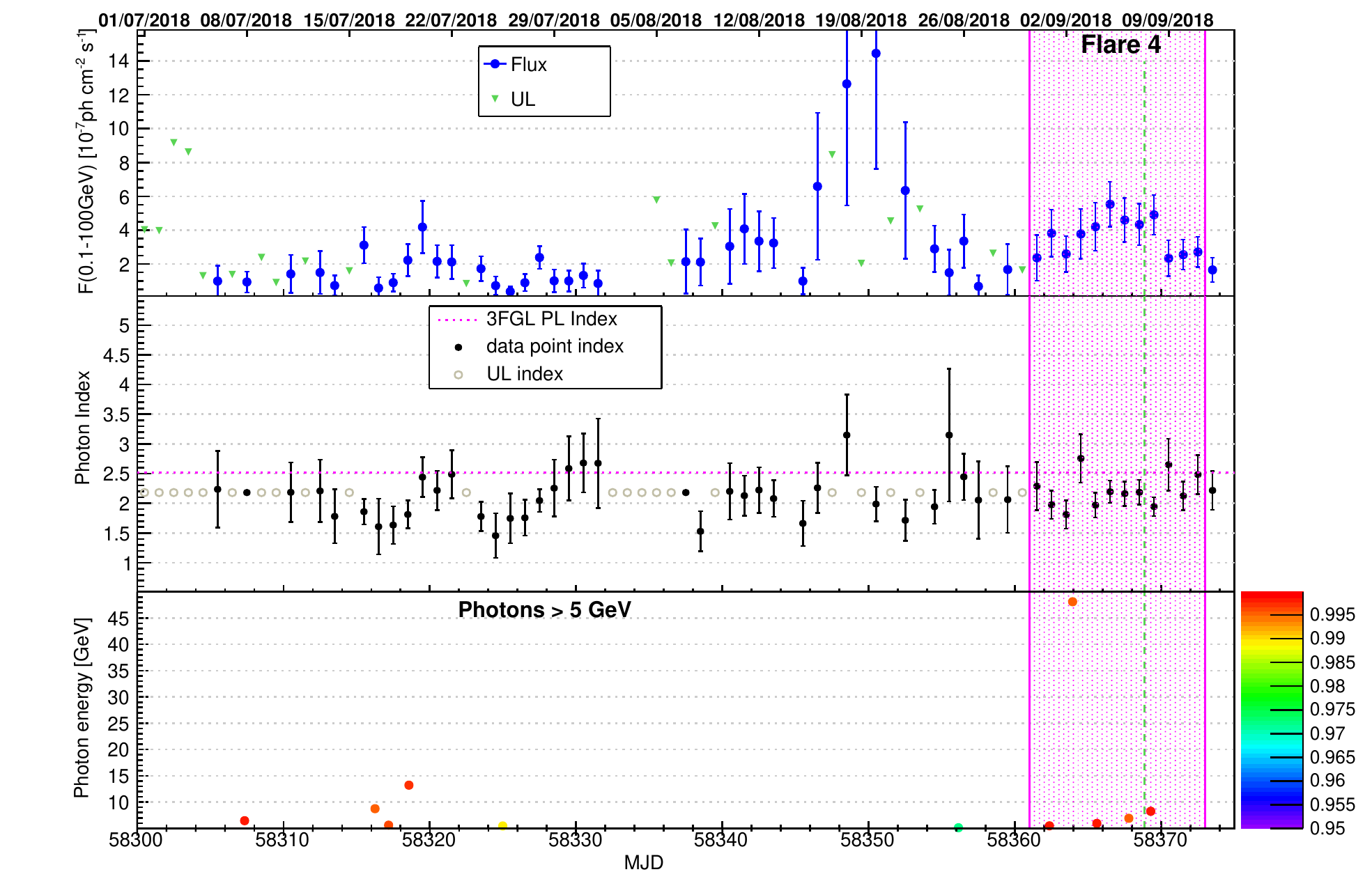} % updated 18 July 2022
\end{center}
\caption{\label{Fig:Figure_9} {\it Fermi}-LAT light-curve, photon index and high-energy photons of NVSS J141922-083830 from 1 July to 14 September 2018 (MJD 58300 to 58375), encompassing the fourth flaring period (highlighted with a magenta shaded area). The time stamp of the MASTER higher detected flux (7 September 2018 / MJD 58368.8545 at B=14.04) is indicated by a vertical green dashed line. Upper panel: {\it Fermi}-LAT light-curve in the 100~MeV--100~GeV energy range, for a one-day binning. Upper limits are represented in green. 
Second panel: power law photon index as optimised by the likelihood analysis for each bin of the {\it Fermi}-LAT light-curve. In light grey are indicated the values used for computing the upper limits (average values). Bottom panel: highest energy "{\sc ultraclean}" photons candidates $>$ 5 GeV. The colour code labeled from 0.95 to 1.00 on the right indicates the probability for each event to be associated to \NVSS.}
\label{Figure-10}
\end{figure*}
%%%%%%%%%%%%%%%%%%%%%%%%%%%%%%%%%%%%%%%%%%%%%%%%%%%%%%%%%%%%%%%%%%%%%%%%%

%%%%%%%%%%%%%%%%%%%%%%%%%%%%%%%%%%%%%%%%%%%%%%%%%%%%%%%%%%%%%%%%%%%%%%%%%%%%
%%%%%%%%%%%%%%%%%%%%%%%%%%%%%%%%%%%%%%%%%%%%%%%%%%%%%%%%%%%%%%%%%%%%%%%%%%%%
\section{{\it Fermi}-LAT observations and analysis}\label{Fermi}
%%%%%%%%%%%%%%%%%%%%%%%%%%%%%%%%%%%%%%%%%%%%%%%%%%%%%%%%%%%%%%%%%%%%%%%%%

We collected data from NVSS J141922-083830 from the beginning of the {\it Fermi}-LAT mission until November 2018. We identified and studied four gamma-ray flares in the 2014--2018 period, labeled as "Flare 1", "Flare 2", "Flare 3" and "Flare 4" (Table \ref{tab:five_phases}). The MASTER-Net flares, that we reported in Section \ref{sec:MASTER}, were identified at the following dates:
\begin{itemize}
    \item 16 May 2012 (MJD 56064);
    \item 24 March 2014 (MJD 56741, during "Flare 1");
    \item 21 February 2015 (MJD 57074.988, during "Flare 3");
    \item 15 July 2018 to 15 September 2018 (MJD 58315–58377, during "Flare~4").
\end{itemize}

We selected photons in the 100~MeV--100~GeV range. We used the Pass~8 (DR2) dataset \citep{Atwood2013} and the software package --- known as "{\it Fermi} Science Tools" (version v10r0p5)\footnote{\url{http://fermi.gsfc.nasa.gov/ssc/data/analysis/software/}}. We selected {\sc ``SOURCE''} events within a 15$^{\circ}$ radius region (region of interest --- or ROI), centered at the position of NVSS J141922-083830. The {\sc ``P8R2\_SOURCE\_V6''} set of instrument response functions (IRFs) was used. Events which triggered the telescope with a zenith angle $\theta_z>90^{\circ}$ were removed in order to avoid contamination from the Earth limb radiation. The signal was extracted using an unbinned likelihood method, coded in the {\it gtlike/pyLikelihood} tool, part of the {\it Fermi} {\it Science Tools}.
The Galactic diffuse emission and the isotropic backgrounds were modelled by the \texttt{gll\_iem\_v06.fits} and the \texttt{iso\_source\_v06.txt} templates, respectively, provided by the {\it Fermi}-LAT Collaboration. Since the likelihood analysis requires the fitting of the spectra of the sources within a certain distance of the source of interest, in order to re-associate each candidate photon of the ROI to its origin, we modelled all the point sources of the 3FGL \citep{3FGL} located in the ROI and in an additional surrounding 10$^{\circ}$ wide annulus (called ``Source Region''). In this model, spectral parameters were kept free (at least partially) for six point sources in the ROI, each one with a detection significance $> 10 \sigma$ in the 3FGL (four years of data). The normalisation parameters of both the isotropic and Galactic models were also kept free. \NVSS was modelled with a power law function for narrow time domain (light-curves) and narrow energy range (SED bins) analysis. (No extended source was found from the 3FGL in this region, apart the Galactic diffuse emission.)

Whenever the test statistics or the number of predicted counts of the source is below 3, for each time or spectral bin, an upper limit is plotted. We computed 95\% confidence upper limits using the \texttt{UpperLimits} class from the Python version of the {\it Fermi} Science Tools, using the output of the actual unbinned likelihood analysis of the corresponding bin\footnote{\url{https://fermi.gsfc.nasa.gov/ssc/data/analysis/scitools/upper_limits.html}}, having fixed the value of the power law spectral indices of the source of interest to their average value over periods similar to those displayed on the light-curves in Figures \ref{Figure-7}, \ref{Figure-8} and \ref{Fig:Figure_9}.

The estimated systematic uncertainty in the effective area is 5\% in the 100~MeV--100~GeV range. The energy resolution ($\Delta E/E$, at 68\% containment) is 20\% at 100 MeV, and between 6 and 10\% over the 1--500 GeV range\footnote{ \url{http://fermi.gsfc.nasa.gov/ssc/data/analysis/LAT\_caveats.html};\\ \url{http://www.slac.stanford.edu/exp/glast/groups/canda/lat\_Performance.htm}}.

%%%%%%%%%%%%%%%%%%%%%%%%%%%%%%%%%%%%%%%%%%%%%%%%%%%%%%%%%%%%%%%%%%%%%%
\subsection{Light-curves in the 100 MeV--100 GeV range} \label{sec:LCs}
%%%%%%%%%%%%%%%%%%%%%%%%%%%%%%%%%%%%%%%%%%%%%%%%%%%%%%%%%%%%%%%%%%%%%%

We produced an 8-year (2008--2016) light-curve of the data we collected from NVSS J141922-083830 at the early stages of our work on this source, in order to search for significant gamma-ray outbursts, using a 3-day time binning. We identified three such events, now labeled as "Flare~1", "Flare~2" and "Flare~3", and also the long "Quiescent" period, all of which were mentioned earlier (see Table~\ref{tab:five_phases}).
Since the source was without significant gamma-ray flares for the first five years of the {\it Fermi} mission, we used this five-year period to represent the quiescent state of the source to which we compared its flux levels and the spectral shapes of the three outburst events\footnote{Though a few minor optical flares were detected during this period (see Figure~\ref{Figure-2:longterm-phot}), we do not consider the possible contamination of these to the quiescent state to be significant, given the long time range of the whole 2008--2013 period.}. We reported part of this preliminary work in \citet{2016sf2a.conf...93B}.

After the detection of the main peak of "Flare~4" by MASTER-Net, we extended our data analysis up to the end of 2018. We present a three-day binned light-curve of the 2013--2018 period that encompasses the four flaring events in Figure \ref{Figure-7}. We chose a three-day binning as a compromise between getting enough statistically significant data points and a fair scanning of the time variability pattern of the source flux. The MASTER-Net optical light-curve is presented on the top panel (sub-range of Figure \ref{Figure-2:longterm-phot}).
On the second panel the {\it Fermi}-LAT light-curve is shown. The four flaring periods are highlighted with magenta shaded regions. Due to the scarce sampling in the optical, we did not perform correlation studies between both bands, but we can see that optical flares (whenever observations were possible) are coincident with gamma-ray flares within the highlighted range (Flares 1, 3 and 4).

In order to investigate the variability pattern of the source in more detail, we also produced one-day binned light-curves. We show these light-curves in the upper panels of Figure \ref{Figure-8} (for Flares 1, 2 and 3) and Figure \ref{Fig:Figure_9} (for Flare~4). In the middle panels are shown the spectral photon indices $\Gamma$. We note that the hardening of the spectrum is significant during parts of Flares 2, 3 and 4 (often with $\Gamma \lesssim 2$), when the source is brighter, which is a trend frequently observed for bright {\it Fermi}-LAT FSRQs and referred as the ``harder when brighter'' pattern. The value of $\Gamma$ becomes lower when the source is brighter. (We note a time gap during MJD 56681--56690 in Figures \ref{Figure-8} and \ref{Fig:Figure_9}, as {\it Fermi}-LAT was mainly operating in pointing mode on other sky regions\footnote{\url{https://fermi.gsfc.nasa.gov/ssc/observations/timeline/posting/ao6/}}.)

%%%%%%%%%%%%%%%%%%%%%%%%%%%%%%%%%%%%%%%%%%%%%%%%%%%%%%%%%%%%%%%%%%%%%%%%%%%%
%%%%%%%%%%%%%%%%%%%%%%%%%%%%%%%%%%%%%%%%%%%%%%%%%%%%%%%%%%%%%%%%%%%%%%%%%%%%
\begin{table*}
  \begin{center}
  \caption{Spectral parameters obtained for NVSS J141922-083830 from the likelihood analysis of the {\it Fermi}-LAT data ($>$ 100 MeV). This is presented for the five periods mentioned in Table \ref{tab:five_phases}, when the source was modeled with a PL.}
  \label{tab:Fluxes_PL}
  \begin{tabular}{cccccccc}
    \hline
    Phase     & $N_0$      & $\Gamma$         & $E_{0}$ & $N_{\rm pred}$ & TS & F(E $>$ 100 MeV) & -ln(Likelihood)\\
              & ($10^{-12}$~cm$^{-2}$~s$^{-1}$~MeV$^{-1}$) &               & (MeV)  &           &     & (10$^{-7}$~ph~cm$^{-2}$~s$^{-1}$) & \\
    \hline
    %Quiescent & 1.50 $\pm$ 0.10 & 2.28 $\pm$ 0.05 & 981.7 & 1950.4 & 443.4 & 0.21 $\pm$ 0.02 & 8672025.5\\
    Quiescent & 1.49 $\pm$ 0.10 & 2.28 $\pm$ 0.05 & 981.7 & 1936.2 & 441.3 & 0.21 $\pm$ 0.02 & 8672025.6\\
    \hline
    %Flare 1   &	9.67 $\pm$ 0.88 & 2.30 $\pm$ 0.07 & 981.7 & 532.4  & 445.9 & 1.42 $\pm$ 0.14 & 344647.8\\
    Flare 1   &	9.69 $\pm$ 0.88 & 2.30 $\pm$ 0.07 & 981.7 & 532.1  & 444.4 & 1.42 $\pm$ 0.14 & 344647.8\\
    \hline
    %\st{Flare 2}   &  \st{32.9 $\pm$ 2.31} &  \st{1.98 $\pm$ 0.05} &  \st{981.7} &  \st{498.9}  &  \st{767.4} &  \st{3.09 $\pm$ 0.27} &  \st{180626.1}\\
    %{\color{magenta}
    %Flare 2   & 32.8 $\pm$ 2.30 & 1.98 $\pm$ 0.05 & 981.7 & 495.8  & 765.5 & 3.07 $\pm$ 0.27 & 180628.2\\
    Flare 2   & 32.67 $\pm$ 2.28 & 1.98 $\pm$ 0.05 & 981.7 & 493.1  & 765.6 & 3.05 $\pm$ 0.26 & 180628.2\\
    \hline
    %Flare 3   &	29.3 $\pm$ 3.46 & 2.09 $\pm$ 0.10 & 981.7 & 187.9  & 288.1 & 3.20 $\pm$ 0.46 & 55330.2\\
    %Flare 3 till MJD 57090 & 19.30 $\pm$ 2.12 &   2.18 $\pm$  0.09 & 981.7 &  260.0 & 317.4 & 2.36 $\pm$ 0.31 & 102207.0\\
    Flare 3 & 19.29 $\pm$ 2.11 &   2.18 $\pm$  0.09 & 981.7 &  259.9 & 317.6 & 2.36 $\pm$ 0.29 & 102207.0\\
    \hline
    %Flare 4   &	30.3 $\pm$ 2.77 & 2.13 $\pm$ 0.07 & 981.7 & 348.0  & 548.7 & 3.50 $\pm$ 0.34 & 84669.5\\
    Flare 4   &	30.30 $\pm$ 2.77 & 2.13 $\pm$ 0.07 & 981.7 & 348.0  & 548.7 & 3.50 $\pm$ 0.34 & 84669.5\\
    \hline
  \end{tabular}
  \end{center}
\end{table*}
%%%%%%%%%%%%%%%%%%%%%%%%%%%%%%%%%%%%%%%%%%%%%%%%%%%%%%%%%%%%%%%%%%%%%%%%%%%%
\begin{table*}
  \begin{center}
  \caption{Spectral parameters obtained for NVSS J141922-083830 from the likelihood analysis of the {\it Fermi}-LAT data ($>$ 100 MeV). This is presented for the five periods mentioned in Table \ref{tab:five_phases}, when the source was modeled with a LP.}
  \label{tab:Fluxes_LP}
  \begin{tabular}{ccccccccc}
    \hline
    Phase     & $N_0$    & $\alpha$  & $\beta$       & $E_{0}$ & $N_{\rm pred}$ & TS & F(E $>$ 100 MeV) & -ln(Likelihood)\\
              & ($10^{-12}$~cm$^{-2}$~s$^{-1}$~MeV$^{-1}$) &   &            & (MeV)  &           &     & {\footnotesize(10$^{-7}$~ph~cm$^{-2}$~s$^{-1}$)} & \\
    \hline
    Quiescent &  1.70 $\pm$ 0.13 & 2.27 $\pm$ 0.06 &  0.095 $\pm$ 0.042 & 981.7 & 1775.0 & 446.7 & 0.18 $\pm$ 0.02 & 8672022.3\\
    \hline
    Flare 1   &	10.43 $\pm$ 1.15 & 2.32 $\pm$ 0.08 &  0.058 $\pm$ 0.055 & 981.7 &  516.4 & 444.9 & 1.34 $\pm$ 0.16  & 344647.1\\
    \hline
    %\st{Flare 2}   &  \st{38.78 $\pm$ 3.37} &  \st{1.94 $\pm$ 0.067} &   \st{0.115 $\pm$ 0.043} &  \st{981.7} &   \st{461.0} &  \st{770.0} &  \st{2.64 $\pm$ 0.29}  &  \st{180650.5}\\
    %\textcolor{magenta}{
    Flare 2 & 38.90 $\pm$ 3.38 & 1.94 $\pm$ 0.07 &  0.117 $\pm$ 0.043 & 981.7 &  460.3 & 772.7 & 2.63 $\pm$ 0.29  & 180623.4\\
    \hline
    %Flare 3   &	30.14 $\pm$ 4.42 & 2.09 $\pm$ 0.100 &  0.018 $\pm$ 0.059 & 981.7 &  185.5 & 287.7 & 3.12 $\pm$ 0.52  &  55330.2\\
    Flare 3  & 19.30 $\pm$  2.12 &   2.18 $\pm$  0.09 & $\sim 0$  & 981.7 & 260.0 & 317.5 & 2.37 $\pm$ 0.31 &  102207.0\\
    \hline
    Flare 4   &	32.62 $\pm$ 3.62 & 2.15 $\pm$ 0.08 &  0.051 $\pm$ 0.048 & 981.7 &  340.9 & 547.9 & 3.34 $\pm$ 0.37  &  84668.9\\
    \hline
  \end{tabular}
  \end{center}
\end{table*}
%%%%%%%%%%%%%%%%%%%%%%%%%%%%%%%%%%%%%%%%%%%%%%%%%%%%%%%%%%%%%%%%%%%%%%%%%%%%

In the lower panels of Figures \ref{Figure-7}, \ref{Figure-8} and \ref{Fig:Figure_9} high energy (HE) photons ($>$ 5~GeV) are shown. These photons were obtained using the \texttt{gtsrcprob} tool\footnote{\url{https://raw.githubusercontent.com/fermi-lat/fermitools-fhelp/master/gtsrcprob.txt}}, within a ROI=0.75$^\circ$. This size of the ROI corresponds to the 68\% containment angle of the acceptance weighted PSF at about 1.2~GeV\footnote{\url{https://www.slac.stanford.edu/exp/glast/groups/canda/archive/pass8v6/lat_Performance.htm}}.
We used the {\sc ULTRACLEAN} class of events, where counts have a higher probability to be real photons, compared to the {\sc source} class events we used to obtain the light-curves and SEDs. The probability level shown on the figure corresponds to the probability that the candidate photons come from our source of interest. In the one-day binned light-curves, the probability levels of the high-energy photons may differ from those of Figure \ref{Figure-7} as there were optimised by using the spectral parameters from the one-day binned likelihood analysis instead of the three-day one. HE photons were mainly observed during Flares 2, 3 and 4, as is often expected during flaring activities when the spectrum of the source becomes harder.

We determined the following highest gamma-ray fluxes of every flaring episode (one day averaged), respectively:
\begin{itemize}
    \item on 12 Mar 2014 (MJD 56728): $(7.02 \pm 2.36) \times 10^{-7}$ph~cm$^{-2}$s$^{-1}$;
    \item on 24 Oct 2014 (MJD 56954): $(7.57 \pm 1.83) \times 10^{-7}$ ph~cm$^{-2}$s$^{-1}$;
    \item on 23 Feb 2015 (MJD 57076): $(5.92 \pm 1.97) \times 10^{-7}$ ph~cm$^{-2}$s$^{-1}$;
    \item on 5 Sep 2018 (MJD 58366): $(5.53 \pm 1.33) \times 10^{-7}$ ph~cm$^{-2}$s$^{-1}$.
\end{itemize}
(Higher fluxes preceding this latter flare, during MJD 58346--58353, could also be reported. However, the large error bars of these data points, that may indicate a low exposure, led us to discard these values -- see Figure \ref{Fig:Figure_9}).

We also observed that the gamma-ray flux from \NVSS did not return down to that of the quiescent state during the long period in between Flare~2 and Flare~4 (from October 2014 to September 2018). Several minor episodes of activity occurred and were often accompanied by the emission of high-energy photons (Figure \ref{Figure-7}).\\

The highest energy photons detected were: at 33.8~GeV on 10 February 2015 (MJD 57063.9701, prob=0.9787) at the start/before Flare~3; at 39.2 GeV on 01 April 2015 (MJD 57113.8687, prob=0.9985), a few days after the end of Flare~3 and before a short secondary outburst; at 48.1 GeV on 2 September 2018 (MJD 58363.9374, prob=0.996243) at the beginning of Flare~4. (These probabilities {\it prob} cited above were calculated through the one-day binned likelihood analysis.) However, given the short duration of these flares and the limited statistics of high-energy photons, it remains difficult to establish a precise pattern on the arrival time of the photons during a flare.\\

We utilized the {\it Gatspy} implementation of the Lomb-Scargle method \citep{2015ApJ...812...18V} to derive a power spectrum of the 3-day binned {\it Fermi}-LAT light curve. No significant peaks were seen.

%%%%%%%%%%%%%%%%%%%%%%%%%%%%%%%%%%%%%%%%%%%%%%%%%%%%%%%%%%%%%%%%%%%%%%
\subsection{Time-resolved SEDs} \label{sec:SEDs}
%%%%%%%%%%%%%%%%%%%%%%%%%%%%%%%%%%%%%%%%%%%%%%%%%%%%%%%%%%%%%%%%%%%%%%

Studies of the quiescent and flaring activities were also performed in the spectral domain, using the unbinned likelihood analysis. We first processed the analysis of each of the five periods (defined in Table~\ref{tab:five_phases}), and obtained the spectral parameters of the best fit models, using both a power law (PL) and a log-parabolic (LP) function for \NVSS (Tables \ref{tab:Fluxes_PL} and \ref{tab:Fluxes_LP}). %

%The result of  $TS_{\rm curv}$ for each of the five periods was smaller than $1.5 \times 10^{-5}$, ruling out any significant curvature of the spectrum.}\\

%%%%%%%%%%%%%%%%%%%%%%%%%%%%%%%%%%%%%%%%%%%%%%%%%%%%%%%%%%%%%%%%%%%%%%%%%%%%
\begin{figure*}
  \begin{center}
    \includegraphics[width=0.75\textwidth]{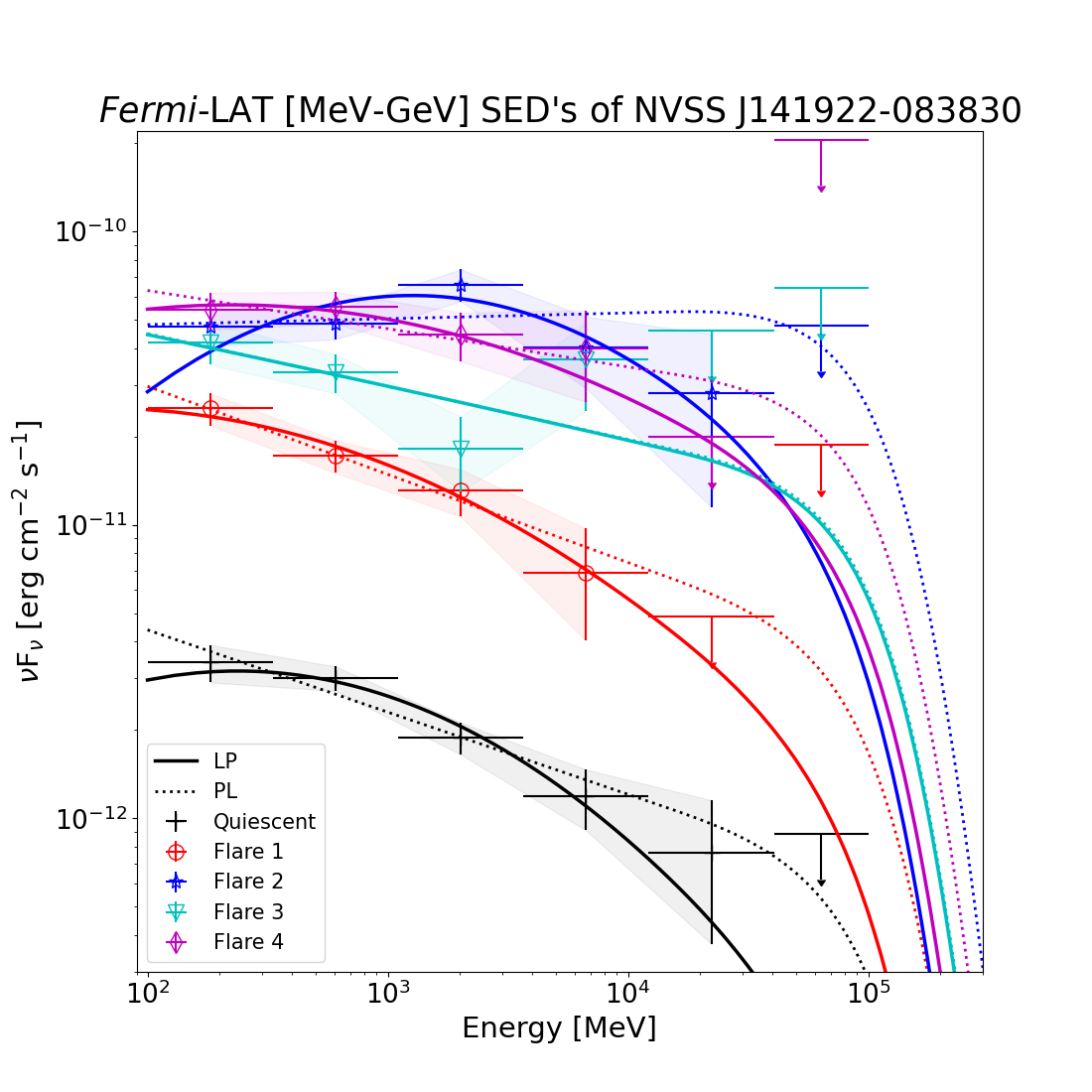} % Figure 10.
  \end{center}
  \caption{\label{Fig:SEDs} {\it Fermi}-LAT SEDs of NVSS J141922-083830, for the quiescent phase, Flare 1, Flare 2, Flare 3 and Flare 4, between 100 MeV and 100 GeV. The dotted and solid lines correspond to respectively the PL and LP fits of the data points, using the parameters obtained by the unbinned likelihood analysis as given in Table \ref{tab:Fluxes_PL} and \ref{tab:Fluxes_LP}. The shaded areas were drawn from the uncertainties on the individual data points.}
  %\label{fig:fermilatSEDs}
\end{figure*}

An optimised model for each period was used to compute an SED with six bins in energy in the 100~MeV--100~GeV energy range. After the optimisation of the spectral parameters over the whole energy range corresponding to each period (shown in Tables \ref{tab:Fluxes_PL} and \ref{tab:Fluxes_LP}), each SED data point was obtained by running again the unbinned likelihood algorithm. This time, in addition to the previous model setup with a limited number of free parameters, we froze all spectral parameters that define the shape of the spectrum to their optimized values, even those of bright sources. The five SEDs are shown in Figure \ref{Fig:SEDs}.

Flare 1 was significantly weaker than the others, Flares 2, 3 and 4 being at a similar flux level. Our results also show significant hardening during Flares~2  ($\Gamma \simeq 2.0$) and indications of hardening during Flares~3 and 4 ($\Gamma \simeq 2.0-2.1$), but not during Flare 1, where $\Gamma \simeq 2.3,$ as during the Quiescent phase (and in the 3FGL). This \textbf{matches} the "harder when brighter" trend which is often observed during major flares in FSRQs \citep{2010ApJ...710.1271A}. 

We fitted the SED data points by using both the PL and LP functions, whose parameters were computed through the unbinned likelihood analysis, respectively (Tables \ref{tab:Fluxes_PL} and \ref{tab:Fluxes_LP}), and by adding the exponential factor $e^{-\tau(E)}$ to account for absorption by the extragalactic background light (EBL). The function $\tau(E)$ represents the EBL optical depth  modelled by \citet{Finke10}, corresponding to a redshift z = 0.90. Hence we produced the fitted SEDs of Figure~\ref{Fig:SEDs} by using the two following equations, respectively:
\begin{equation}
\left(\frac{dN}{dE} (E) \right)_{PL}  = N_0 \left(\frac{E}{E_0}\right)^{-\Gamma} e^{-\tau(E)}; \hspace{1.4cm}
\end{equation}

\begin{equation}
\left( \frac{dN}{dE}(E) \right)_{LP} = N_0 \left(\frac{E}{E_0}\right)^{\alpha-\beta~ln(E/E_0 )} e^{-\tau(E)},
\end{equation}
where $E_0$ = 981.713 MeV is the pivot energy of the \NVSS spectrum as reported in the 3FGL catalogue from the PL fit, and $\tau(E)$ is the opacity to the extragalactic background light\footnote{The values of $\tau(E)$ at 20, 50 and 100~GeV are equal to 0.005, 0.166 and 0.798, respectively.}.
({Note:} we needed to assume that the fit parameters found with the likelihood analysis -- without EBL modelling -- remained the best estimated ones even in the case where the factor $e^{-\tau(E)}$ is included in the fit functions. We do consider this approximation acceptable, since the departure from the PL fit is only observed from $\sim$20 GeV where the photon statistics is low.)
%%%%%%%%%%%%%%%%%%%%%%%%%%%%%%%%%%%%%%%%%%%%
\begin{table}
  \begin{center}
  \caption{Values of $TS_{\rm curv}$ computed with Equation \ref{eq:TScurv} and values of $\chi^2/ndf$ for fitted data points, for the five observation periods.} \label{tab:TScurv}
\begin{tabular}{c|c|c|c}
\hline
          &   & \multicolumn{2}{|c|}{$\chi^2/ndf$ (data point fits)}\\
 Period   & $TS_{\rm curv}$    &    \multicolumn{2}{|c|}{-------------------------------------}\\                           
          &   & $PLe^{-\tau(E)}$  & $LPe^{-\tau(E)}$  \\
\hline
%Quiescent &   6.4  & 0.18  & 0.17\\
%Flare 1   &   1.4  & 0.077 & 0.082\\
%\st{Flare 2}   & \st{-48.8}  & \st{1.37}  & \st{0.42}\\
%Flare 2   &   9.5  &  1.39 & 0.42\\
%Flare 3   &   0.0  & 1.11  & 1.13\\
%Flare 3 till MJD 57090 & 0.0 & 0.86 & 0.86\\
%Flare 4   &   1.2  & 0.21  & 0.11\\
Quiescent &   6.5  & 0.19  & 0.17\\
Flare 1   &   1.3  & 0.080 & 0.082\\
Flare 2   &   9.6  &  1.39 & 0.42\\
Flare 3   &   0.0  & 0.86  & 0.86\\
Flare 4   &   1.2  & 0.21  & 0.11\\
\hline
\end{tabular}
  \end{center}
\end{table}
%%%%%%%%%%%%%%%%%%%%%%%%%%%%%%%%%%%%%%%%%%%%

We calculated the curvature spectrum significance by checking the compatibility of the LP function with the PL function, using:

%\begin{linenomath*}
\begin{equation} \label{eq:TScurv}
TS_{\rm curv} = 2 \times [ ln (\mathcal{L}_{LP} ) - ln (\mathcal{L}_{PL} )],
\end{equation}
%\end{linenomath*}
where $ln(\mathcal{L}_{LP})$ and $ln(\mathcal{L}_{PL})$ are the natural logarithm of the maximum likelihood obtained with the LP and PL models, respectively, and reported in Tables \ref{tab:Fluxes_PL} and \ref{tab:Fluxes_LP}. The curvature is considered to be significant if $TS_{\rm curv} \ge 4$ \citep{4FGL_DR3}. Results are given in Table \ref{tab:TScurv}. Significant curvature is not observed for Phases 1, 3 and 4, but is observed for the Quiescent phase ($TS_{\rm curv} = 6.5$) and Flare~2 ($TS_{\rm curv} = 9.6$).

Fits on the binned SED data points allow for further investigation on the departure from either the PL or LP function. We find most of the computed $\chi^2/ndf$ values for each fit to be very small or $\lesssim 1.1$, except for the PL fit of Flare 2, where $\chi^2/ndf \simeq 1.4$. We also note a higher curvature during this flare compared to the quiescent state and the other flaring periods ($\beta = 0.117 \pm 0.043$ --- see Table~\ref{tab:Fluxes_LP}.)

%\begin{table}
%  \begin{center}
%  \caption{Gamma-ray isotropic luminosities during the five observation periods. These luminosities were derived using the integrated flux values in the 100 MeV--12 GeV range, found using a LP spectral fit (Equation~\ref{eq:L}.)}
%  \label{tab:gamma_lum}
%  \begin{tabular}{ccc}
%\hline
%  \multirow{}{}{} & Period & L (10$^{47}$ erg~s$^{-1}$)\\
%\hline
%\multirow{} & Quiescent & 0.5\\
%\multirow{$H_0$=67.4 k$^{-1}$ %Mpc$^{-1}$} & Flare 1 & 3.2 \\
%\multirow{} & Flare 2 & 10.2 \\
%\multirow{$d_L$=19.570 Gly} & Flare %3 & 6.2 \\ 
%\multirow{}{}{} & Flare 4 & 9.5 \\
%\hline
%\multirow{}{} & Quiescent & 0.4\\
%\multirow{}{$H_0$=73.2 km s$^{-1}$ %Mpc$^{-1}$} & Flare 1 & 2.7 \\
%\multirow{} & Flare 2 & 8.6 \\
%\multirow{}{$d_L$=18.019 Gly} & %Flare 3 & 5.3 \\ 
%\multirow{}& Flare 4 & 8.1 \\
%\hline
%  \end{tabular}
%  \end{center}
%\end{table}

\begin{table}
  \begin{center}
  \caption{Gamma-ray isotropic luminosities during the five observation periods. These luminosities were derived using the integrated flux values in the 100 MeV--12 GeV range, found using a LP spectral fit (Equation~\ref{eq:L}.)}
  \label{tab:gamma_lum}
  \begin{tabular}{ccc}
\hline
  \multirow{3}{*} & Period & L (10$^{47}$ erg~s$^{-1}$)\\
\hline
\multirow{3}{*}{}& Quiescent & 0.5\\
\multirow{3}{*}{$H_0$=67.4 k$^{-1}$ Mpc$^{-1}$} & Flare 1 & 3.2 \\
 & Flare 2 & 10.2 \\
\multirow{3}{*}{$d_L$=19.570 Gly} & Flare 3 & 6.2 \\ 
 & Flare 4 & 9.5 \\
\hline
\multirow{3}{*} & Quiescent & 0.4\\
\multirow{3}{*}{$H_0$=73.2 km s$^{-1}$ Mpc$^{-1}$} & Flare 1 & 2.7 \\
& Flare 2 & 8.6 \\
\multirow{3}{*}{$d_L$=18.019 Gly} & Flare 3 & 5.3 \\ 
&Flare 4 & 8.1 \\
\hline
  \end{tabular}
  \end{center}
\end{table}
%%%%%%%%%%%%%%%%%%%%%%%%%%%%%%%%%%%%%%%%%%%%%%%%%%%%%%%%%%%%%%%%%%%%%%
\subsection{Gamma-ray luminosities} \label{sec:Lum}
%%%%%%%%%%%%%%%%%%%%%%%%%%%%%%%%%%%%%%%%%%%%%%%%%%%%%%%%%%%%%%%%%%%%%%

Based on the redshift measurement reported in Section~\ref{sec:spectro} (z = 0.903) and the results of the likelihood analysis presented in Table~\ref{tab:Fluxes_LP}, we calculated the integrated apparent luminosity $L$ (isotropic assumption) of the source for the quiescent and flaring periods in the 100~MeV--12~GeV range, using the formula

\begin{equation}
L = 4 \pi d_{L}^{2} \int_{E_1}^{E_2} E \frac{dN(E)}{dE} dE, \label{eq:L}
\end{equation}
where $E$, $E_1$ = 100 MeV and $E_2$ = 12 GeV are the energies in the observer frame, and $d_L$ is the luminosity distance (considering 1ly = $9.461.10^{26}$ cm).

We used the LP spectral shape. The choice of integrating up to 12~GeV was motivated by the fact that we do not get significant signal beyond this value in the SEDs. We use the cosmological parameters $\Omega_m$ = 0.315 and $\Omega_{\lambda}$ = 0.685 \citep{2020A&A...641A...6P}. Due to the uncertainty on the measurement of the Hubble constant, we computed results with both $H_0$=67.4 \citep{2020A&A...641A...6P} and 73.2 km~s$^{-1}$~Mpc$^{-1}$ \citep{2016ApJ...826...56R}. Results are given in Table~\ref{tab:gamma_lum}.

%%%%%%%%%%%%%%%%%%%%%%%%%%%%%%%%%%%%%%%%%%%%%%%%%%%%%%%%%%%%%%%%%%%%%%
\section{Broad-band SED modeling} \label{sec:broadband_SED}
%%%%%%%%%%%%%%%%%%%%%%%%%%%%%%%%%%%%%%%%%%%%%%%%%%%%%%%%%%%%%%%%%%%%%%
\begin{figure*}
%\centering
%\includegraphics[scale=0.6]{NVSS1419SEDfits_3Jul2021.pdf} % previous Flare 2
\includegraphics[scale=0.6]{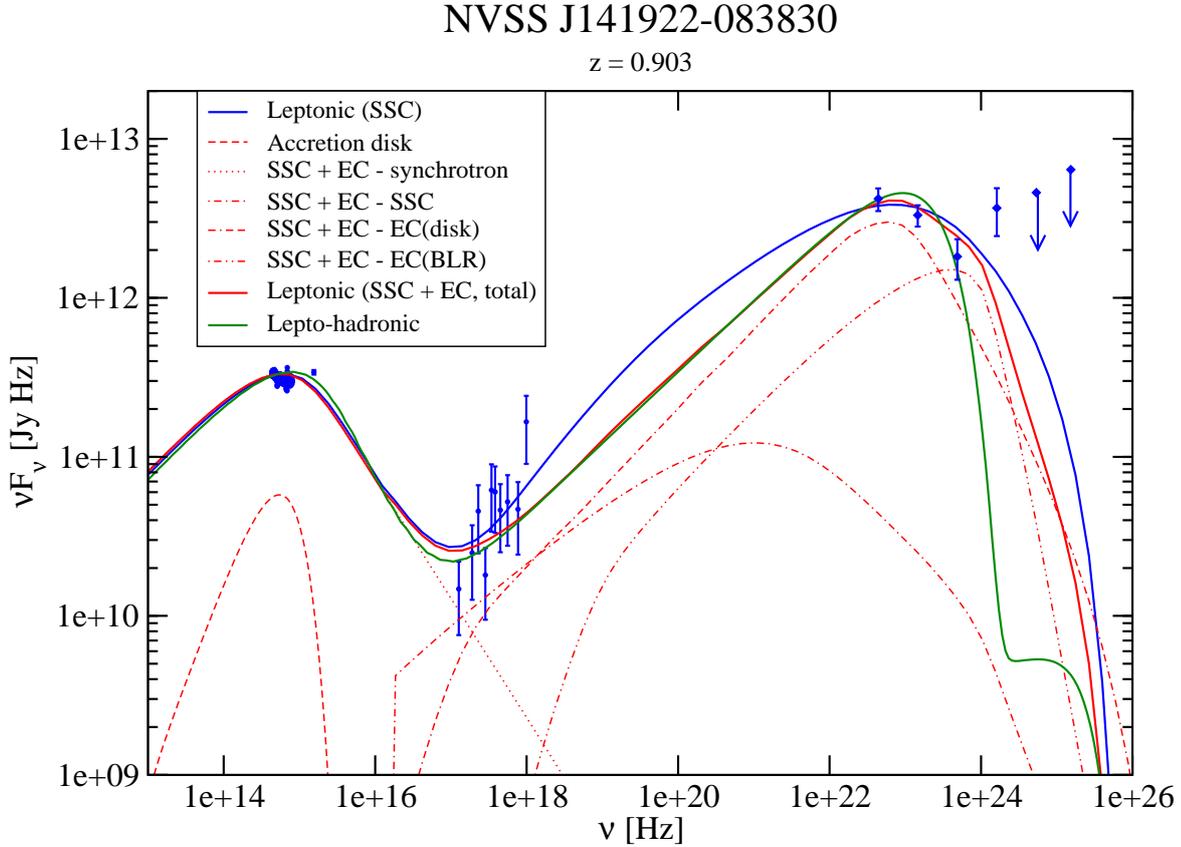} % extended Flare 2
\caption{Multiwavelength SED of \NVSS during the February--March 2015 flare. 
Three steady-state single-zone model fits are presented: Blue: A leptonic 
model with only synchrotron and synchrotron-self-Compton emission; red:
A leptonic model with also external-Compton emission included: dot-dashed
= SSC, dot-dot-dashed = EC on direct accretion-disk photons; dot-dashed-dashed:
EC on BLR emission; green: a lepto-hadronic model, where the $\gamma$-ray
emission is dominated by proton synchrotron. See text and Table \ref{model_parameters} for details and parameters, respectively. }
\label{fig:SEDfits}
\end{figure*}
%%%%%%%%%%%%%%%%%%%%%%%%%%%%%%%%%%%%%%%%%%%%%%%%%%%%%%%%%%%%%%%%%%%%%%
\begin{table*}
  
  %\centering
  \caption{Model parameters for broad-band SED model fits.\label{model_parameters} }
  \begin{center}
  \begin{tabular}{lccc}
    \hline
      Parameter & Leptonic SSC & Leptonic SSC + EC & Lepto-hadronic \\
    \hline
      Minimum electron injection $\gamma$ & $1.1 \times 10^4$ & $1.3 \times 10^3$ & 450 \\
      Maximum electron injection $\gamma$ & $2.0 \times 10^5$ & $2.0 \times 10^5$ & $1.0 \times 10^4$ \\
      Electron injection spectral index   & 3.5 & 3.5 & 4.0 \\
      Escape timescale parameter $\eta_{\rm esc}$ & 10 & 10 & 1 \\
      Magnetic field [G] & 0.2 & 9 & 100 \\
      Radius of emission region [cm] & $2.5 \times 10^{16}$ & $1.5 \times 10^{15}$ & $1.0 \times 10^{16}$ \\
      Bulk Lorentz factor $\Gamma_{\rm jet}$ & 10 & 20 & 20 \\
      Emission region distance from BH [pc] & --- & 0.028 & --- \\
      Black-hole mass [M$_{\odot}$] & --- & $5 \times 10^9$ & --- \\
      Accretion-disk luminosity [erg s$^{-1}$] & --- & $1.0 \times 10^{46}$ & --- \\
      External radiation field T [K] & --- & $4.0 \times 10^4$ & --- \\
      External radiation field u [erg cm$^{-3}$] & --- & $5.0 \times 10^{-2}$ & --- \\
      \hline
      Minimum non-thermal proton energy [GeV] & --- & --- & 1 \\
      Maximum non-thermal proton energy [GeV] & --- & --- & $2.8 \times 10^8$ \\
      Proton injection spectral index & --- & --- & 2.1 \\
      \hline 
      Kinetic luminosity in radiating electrons [erg s$^{-1}$] & $2.8 \times 10^{45}$ & $2.5 \times 10^{44}$ & $2.6 \times 10^{41}$ \\
      Poynting-flux luminosity [erg s$^{-1}$] & $9.4 \times 10^{42}$ & $2.7 \times 10^{44}$ & $1.5 \times 10^{48}$ \\
      Kinetic luminosity in radiating protons [erg s$^{-1}$] & --- & --- & $1.9 \times 10^{47}$ \\
      Magnetization $L_B/L_e$ & $3.3 \times 10^{-3}$ & 1.07 & $5.8 \times 10^6$ \\
      Magnetization $L_B/L_p$ & --- & --- & 7.8 \\
      Minimum variability time scale [h] & 32 & 1.3 & 7.8 \\ 
      \hline
  \end{tabular}
  \end{center}
\end{table*}

In order to provide further constraints on the classification of
\NVSS, we have compiled a quasi-simultaneous multi-wavelength SED
around the time of the February--March 2015 flare
(Flare 3), since this flare occuured at a time when the available multi-wavelength coverage was best. It includes NIR photometry
from the CANICA/NIR camera of the 2.1m telescope of the Guillermo Haro Observatory \citep{Carrasco}, at MJD 57080.5378 (2015-02-27 12:54:25.920 
UTC), Swift UVOT optical/near-UV photometry from the period during MJD 
57070-57090, the Swift XRT X-ray spectrum from MJD 57085.1582 -- 
57085.2836 (2015-03-04) as well as the {\it Fermi}-LAT spectrum from the 
period MJD 57067 -- 57090. The quasi-simultaneous SED is plotted in 
Figure \ref{fig:SEDfits}. It shows the peak of the low-frequency SED component
around optical wavelengths and the high-energy component peaking at sub-GeV
energies, with a strong $\gamma$-ray dominance, the peak of the $\gamma$-ray 
component being about a factor 10 brighter than the low-frequency (synchrotron) 
component. These are SED features that are characteristic of LSP blazars, such
as FSRQs and low-frequency-peaked BL Lac objects. 

For illustrative purposes, we model this SED with the single-zone
leptonic and lepto-hadronic models developed by \cite{Boettcher}. 
These are steady-state models, which is justified for our purposes,
since the data are not exactly simultaneous and the integration times 
for most of these observations are longer than the characteristic dynamical
and radiative cooling time scales for electrons (and protons) emitting in 
the optical through $\gamma$-ray regimes. 

The model is based on a single emission zone, assumed spherical in
the co-moving frame, which is moving with Lorentz factor $\Gamma_{\rm jet}$ along
the jet. A power-law energy distribution of electrons (and protons in the 
case of the lepto-hadronic model) is injected, and an equilibrium between
injection, radiative (and adiabatic for protons in the lepto-hadronic
model) energy losses, and escape from the emission region is calculated
self-consistently to evaluate the final radiation spectrum. The particle escape
time scale is parameterized as a multiple $\eta_{\rm esc} \ge 1$ of the light
crossing time, such that $t_{\rm esc} = \eta_{\rm esc} \, R/c$, where $R$ is 
the radius of the emission region. Synchrotron
emission, Compton scattering (both synchrotron self-Compton [SSC] and external
Compton [EC]), and, for the lepto-hadronic model, photo-pair and photo-pion
production are included. Internal $\gamma\gamma$ absorption and 
corresponding pair injection and cascading are also treated in a 
steady-state fashion, as detailed in \cite{Boettcher}.
%The model SED is corrected for $\gamma\gamma$
%absorption by the extragalactic background light, which is evaluated using 
%the model of \cite{Finke10} for the redshift of $z = 0.903$.
The model SED includes the gamma-gamma absorption by the EBL, which is evaluated using the model of Finke et al. (2010) for the redshift of z = 0.903.
In order to reduce the number of free parameters, the viewing angle is chosen
as $\theta_{\rm obs} = 1/\Gamma_{\rm jet}$. In the lepto-hadronic models, external
photon fields (both for Compton scattering and for photo-hadronic processes)
are not included. 

As is obvious from Figure \ref{fig:SEDfits}, the SED is not very well sampled, so
there will be large parameter degeneracies in our fits. Our model fits are therefore not meant to provide meaningful constraints on the emission-region
parameters, but just to illustrate that it is possible to represent the SED with
standard one-zone emission scenarios. To illustrate this, we attempt three different
fits: (1) the simplest possible leptonic model fit, including only synchrotron
and SSC; (2) a leptonic model fit, including external radiation fields; and (3) 
a lepto-hadronic model. All 3 of these fits are shown in Figure \ref{fig:SEDfits}, and 
the adopted relevant model parameters are listed in Table \ref{model_parameters}. 
We will discuss these three fits individually below. 

(1) The pure SSC model is the one with the fewest free parameters. Thus, even while 
specific parameter values are still degenerate, in order not to over-shoot the 
observed X-ray spectrum, a very large minimum electron Lorentz factor is required.
The large Compton dominance of the source along with the wide separation of the 
synchrotron and SSC peaks also requires a low magnetic field, leading to a strongly
particle-dominated energy budget in the source, far away from equipartition. This
is a common problem and one of the reasons why pure SSC models are generally 
dis-favoured for LSP blazars. 

(2) A leptonic model including external radiation fields is capable of reproducing
the SED of \NVSS\ with significantly more natural parameters, typical of FSRQs \textbf{and many ISP blazars}. In
particular, model parameters can be chosen that correspond to equipartition between
the kinetic energy in electrons and the magnetic field. In our model fit, the  
$\gamma$-ray emission is dominated by EC on direct disk emission, with sub-dominant
contribution from EC on an isotropic (in the AGN rest frame) thermal radiation field 
with parameters that are appropriate to represent BLR emission \citep[see][for a
discussion]{Boettcher}. Due to a very compact emission region, this model allows
for sub-hour variability. 

(3) Our fit using a lepto-hadronic model is based on strongly proton-synchrotron 
dominated high-energy emission. The relativistic electron population is energetically
strongly sub-dominant. The energy carried along the jet is dominated by the strong 
magnetic field, as is typical for hadronic and lepto-hadronic models. 

Based on SED considerations alone, we cannot claim a strong preference for either a
leptonic SSC + EC model or a lepto-hadronic model.

%%%%%%%%%%%%%%%%%%%%%%%%%%%%%%%%%%%%%%%%%%%%%%%%%%%%%%%%%%%%%%%%%%%%%%
\section{Summary and conclusions}

We have presented results of multi-wavelength observations of the blazar candidate NVSS J141922-083830, including optical photometry, polarimetry, and spectroscopy. Four large $\gamma$-ray flares were identified, with characteristic time scales of several days.

The optical spectra obtained with SALT-RSS showed two emission lines, identified as the Mg~{\sc ii} 2798\AA\ and C~{\sc iii]} 1909\AA\ lines, yielding a redshift of $z = 0.903$ for the source.

%implying $p = 13.5 \%$$ and $$\theta = -18.2 ±12.5^\circ$.
Optical polarimetry revealed variable polarization at a level of $\Pi \sim 10$~\%. This provides evidence for the synchrotron origin of the optical emission and further supports the blazar nature of this source. Given a characteristic optical spectral index of $\alpha \sim 1$, a perfectly ordered magnetic field in the optical emission region would lead to a degree of polarization of $\sim 75$~\%. The significantly lower degree of polarization therefore indicates a partially ordered magnetic field.

We report a hardening of the gamma-ray spectrum during the last three flaring periods, with a power-law spectral index $\Gamma = 2.0$--$2.1$, while the average flux was $\approx$ 3 $\times$ 10$^{-7}$ ph~cm$^{-2}$s$^{-1}$. The maximum daily gamma-ray flux level was observed during MJD 56954 at $(7.57\pm1.83)\times10^{-7}$ ph~cm$^{-2}$s$^{-1}$, during Flare 2.

The broad-band SED of NVSS J141922-083830 shows the characteristic double-bump structure with a low-frequency peak in the optical and a high-frequency peak at multi-MeV energies. These features, along with a Compton dominance factor of $\sim 10$ are characteristic of LSP blazars, such as FSRQs, although the synchrotron peak frequency of NVSS J141922-083830 appears to be located in the optical, which is more characteristic of an ISP blazar. However, there is some uncertainty on the position of the synchrotron peak due to the limited amount of data. A single-zone, leptonic model with an external radiation field characteristic of BLR radiation, is able to reproduce the SED with plausible parameters characteristic of FSRQs \citep[e.g.,][]{Ghisellini10,Boettcher} and equipartition between relativistic electrons and the magnetic field. Also a proton-synchrotron-dominated hadronic model can not be ruled out, but requires an extreme energy budget in excess of $10^{48}$~erg~s$^{-1}$.

%Black-hole mass [M$_{\odot}$] & --- & $5 \times 10^9$ 
%Accretion-disk luminosity [erg s$^{-1}$] & --- & $1.0 \times 10^{46}$ 

%%%%%%%%%%%%%%%%%%%%%%%%%%%%%%%%%%%%%%%%%%%%%%%%%%%%%%%%%%%%%%%%%%%%%%

\section*{Acknowledgements}

The \textit{Fermi}-LAT Collaboration acknowledges generous ongoing support
from a number of agencies and institutes that have supported both the
development and the operation of the LAT as well as scientific data analysis.
These include the National Aeronautics and Space Administration and the
Department of Energy in the United States, the Commissariat \`a l'Energie Atomique
and the Centre National de la Recherche Scientifique / Institut National de Physique
Nucl\'eaire et de Physique des Particules in France, the Agenzia Spaziale Italiana
and the Istituto Nazionale di Fisica Nucleare in Italy, the Ministry of Education,
Culture, Sports, Science and Technology (MEXT), High Energy Accelerator Research
Organization (KEK) and Japan Aerospace Exploration Agency (JAXA) in Japan, and
the K.~A.~Wallenberg Foundation, the Swedish Research Council and the
Swedish National Space Board in Sweden.
 
Additional support for science analysis during the operations phase is gratefully acknowledged from the Istituto Nazionale di Astrofisica in Italy and the Centre National d'\'Etudes Spatiales in France.

The SALT spectra were obtained under the programmes 2014-2-DDT-002 (PI: AK) and 2016-2-LSP-001 (PI: DB).

DB, RJB, MB, SR and AK acknowledge support from the National Research Foundation, South Africa and the South African Gamma-ray Astronomy Programme (SA-GAMMA). 

The work of MB is supported through the South African Research Chair Initiative 
of the National Research Foundation\footnote{Any opinion, finding and conclusion or 
recommendation expressed in this material is that of the authors and the NRF does 
not accept any liability in this regard.} and the Department of Science and Innovation 
of South Africa, under SARChI Chair grant No. 64789.

The work of VK was supported by the Ministry of science and higher education of Russian Federation, topic № FEUZ-2020-0038.

MASTER is supported by Lomonosov Moscow State University Development Program. EG is supported by RFBR 19-29-11011.

This work has made use of data from the European Space Agency (ESA)
mission {\it Gaia} (\url{https://www.cosmos.esa.int/gaia}), processed by
the {\it Gaia} Data Processing and Analysis Consortium (DPAC,
\url{https://www.cosmos.esa.int/web/gaia/dpac/consortium}). Funding
for the DPAC has been provided by national institutions, in particular
the institutions participating in the {\it Gaia} Multilateral Agreement.

We thank Abhishek Desai for his reading of the draft and his valuable comments as internal reviewer. We also thank other members of the Fermi LAT Collaboration, Josefa Becerra, Matthew Kerr and Deirdre Horan, for their additional and useful comments. %And we thank Sara Buson, Deirdre Horan and Josefa Becerra, Philippe Bruel and Matthew Kerr for their final comments on the draft, on behalf of the \textit{Fermi}-LAT Collaboration.

\section*{Data availability}
The data underlying this article will be shared on reasonable request to the corresponding authors.

\bibliographystyle{mnras}
\bibliography{references} % if your bibtex file is called example.bib

% Don't change these lines
\bsp	% typesetting comment
\label{lastpage}
\end{document}